# Charge Transfer from Photoexcited Semiconducting Single-Walled Carbon Nanotubes to Wide-Bandgap Wrapping Polymer

Zhuoran Kuang, Felix J. Berger, Jose Luis Pérez Lustres, Nikolaus Wollscheid, Han Li, Jan Lüttgens, Merve Balcı Leinen, Benjamin S. Flavel, Jana Zaumseil, and Tiago Buckup*

**ABSTRACT:** As narrow optical bandgap materials, semiconducting single-walled carbon nanotubes (SWCNTs) are rarely regarded as charge donors in photoinduced charge-transfer (PCT) reactions. However, the unique band structure and unusual exciton dynamics of SWCNTs add more possibilities to the classical PCT mechanism. In this work, we demonstrate PCT from photoexcited semiconducting (6,5) SWCNTs to a wide-bandgap wrapping poly-[(9,9-dioctylfluorenyl-2,7-diyl)-*alt*-(6,6′)-(2,2′-bipyridine)] (PFO−BPy) via femtosecond transient absorption spectroscopy. By monitoring the spectral dynamics of the SWCNT polaron, we show that charge transfer from photoexcited SWCNTs to PFO−BPy can be driven not only by the energetically favorable $E_{33}$ transition but also by the energetically unfavorable $E_{22}$ excitation under high pump fluence. This unusual PCT from narrow-bandgap SWCNTs toward a wide-bandgap polymer originates from the up-converted high-energy excitonic state ($E_{33}$ or higher) that is promoted by the Auger recombination of excitons and charge carriers in SWCNTs. These insights provide new pathways for charge separation in SWCNT-based photodetectors and photovoltaic cells.

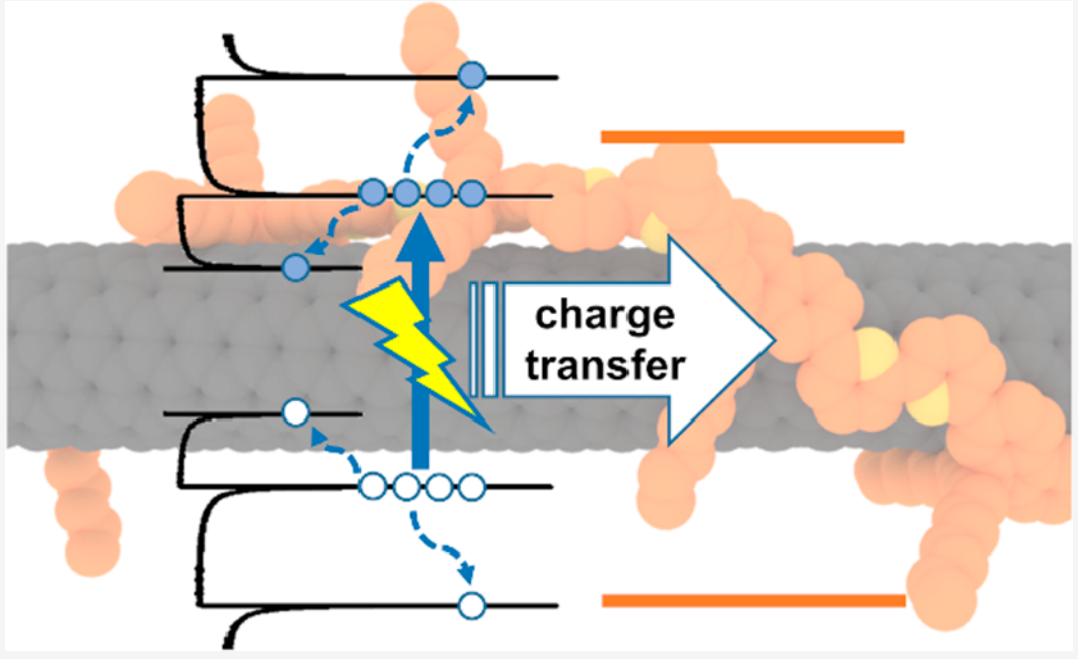

## ■ INTRODUCTION

The highly selective wrapping of semiconducting single-walled carbon nanotubes (SWCNTs) with conjugated polymers, e.g., polyfluorenes and polythiophenes, has enabled monochiral samples with high purity[1−3] to be available for a wide range of optoelectronic applications from field-effect transistors,[4] light-emitting diodes,[5] and electrochromic cells[6] to photovoltaic cells[7,8] and photodiodes.[9] The interaction of nanotubes with their immediate environment, i.e., the wrapping polymer or matrix, with regard to energy and charge transfer depending on the energy level alignment[10,11] is crucial for their functionality and not yet fully understood. The unique electronic structure and complex photophysics of SWCNTs[12] make this interaction highly interesting from a fundamental and application point of view. For photovoltaic cells based on carbon nanotubes, SWCNTs assume the role of donor in photoinduced charge transfer (PCT) only when a semiconductor with high electron affinity (e.g., fullerene derivatives) acts as an acceptor.[13−15] When wrapped by typical semiconducting polymers, the narrow-bandgap nanotubes usually behave as acceptors for both charges and excitation.[16−21]

Various spectroscopic methods, including pump−probe transient absorption (TA) and transient fluorescence spectroscopy, have been applied to explore the excited-state interactions between SWCNTs and the conjugated wrapping polymer.[11,16,22−25] Strong electronic interaction in SWCNT/polymer hybrids has been concluded by studying the modulation effect of the polymer frontier orbital levels on nanotube valence and conduction band energies, as well as newly formed hybridized electronic states between the two components.[22,23,25] The energy transfer observed in SWCNT/polymer composites also points to an electron-exchange mechanism.[11,16] However, only the first optical bandgap in semiconductors is usually taken into account in studies of the classical PCT mechanism. The electronic interplay between the complex band structure of SWCNTs and the wrapping polymer has not been clarified yet.[26,27]

(6,5) SWCNTs are chosen for this study due to their availability as nearly monochiral samples in large amounts after selective dispersion with a polyfluorene−bipyridine copolymer (PFO−BPy).[2] The energy level alignment of (6,5) SWCNTs and PFO−BPy (Figure 1) indicates that the PFO−BPy-wrapped (6,5) SWCNT hybrid system (hereafter referred to as **Hybrid**) constitutes a Type-I heterojunction. When regarding the narrow-bandgap (6,5) SWCNT as a charge donor, the PCT from excited (6,5) SWCNT toward wide-bandgap acceptor PFO−BPy is energetically unfavorable, at least for the $E_{11}$ and $E_{22}$ transitions. However, the established exciton photophysics of SWCNTs suggests that low-energy photon excitation is able to promote populations of high-energy excitonic states via, e.g., Auger recombination of excitons.[30−36] Under this mechanism, even in the presence of ultrafast intersubband relaxation, high-energy excitonic states still have a considerable decay time as that of the $E_{11}$ state, which may favor the PCT from excited SWCNTs to a wide-bandgap acceptor.[35] Here, we analyze this SWCNT-based Type-I heterojunction by femtosecond TA spectroscopy. A comparison with surfactant-dispersed (6,5) SWCNTs in water (hereafter referred to as **SWCNT**), which allows for observations of exciton dynamics without energy or charge transfer,[37−40] is conducted throughout this work. The exciton dynamics of the **Hybrid** in tetrahydrofuran and **SWCNT** in water are investigated upon the $E_{11}$, $E_{22}$, and $E_{33}$ excitation with variable excitation fluences, and PCT products in the **Hybrid** are analyzed. We demonstrate that the PCT from photoexcited (6,5) SWCNT to PFO−BPy is driven by the energetically favorable $E_{33}$ excitation and may be also driven by the energetically unfavorable $E_{22}$ excitation. The latter process occurs via Auger recombination of excitons and charge carriers at high excitation fluences.

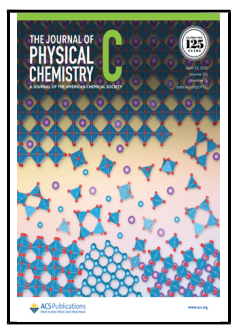

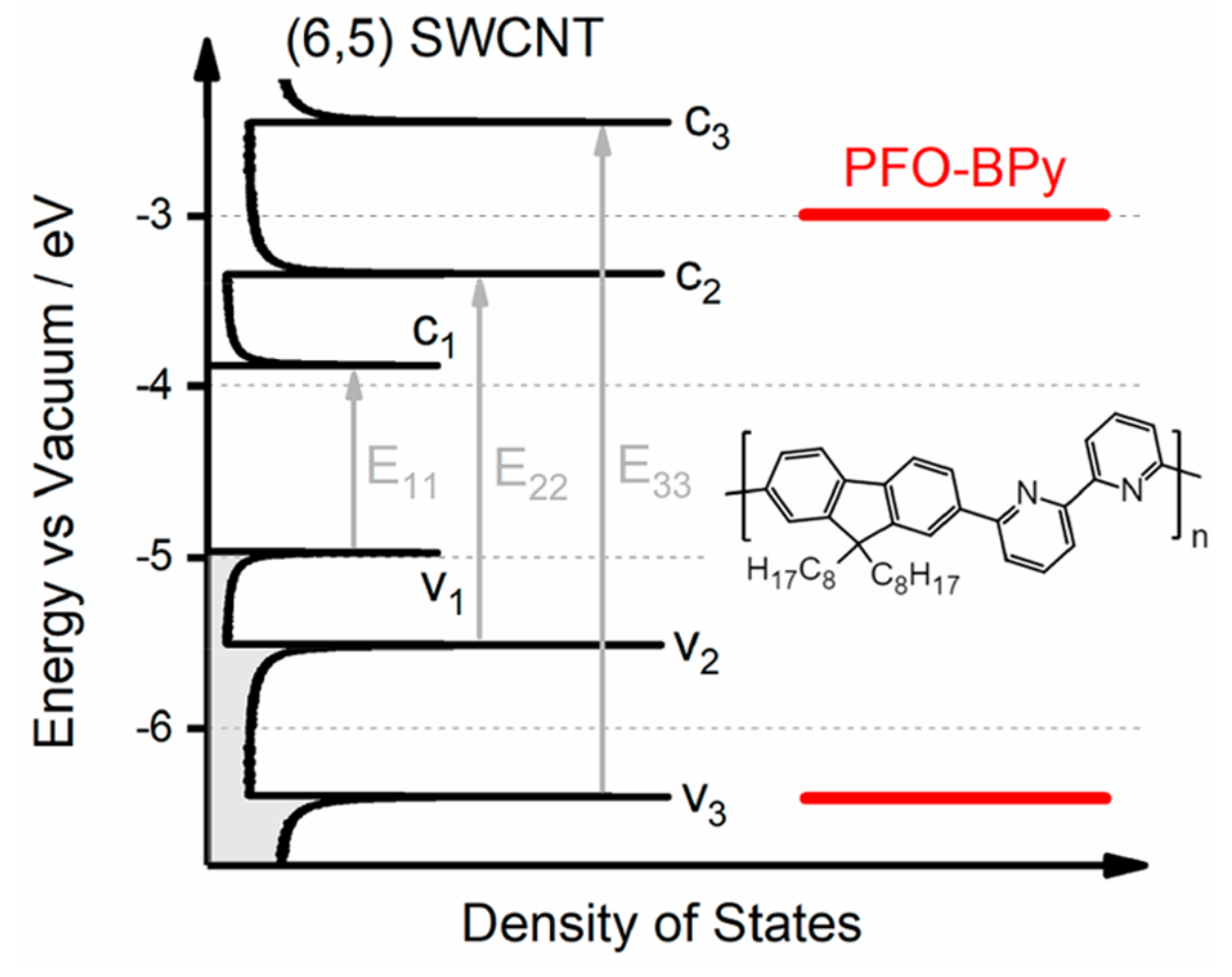

**Figure 1.** Schematic energy level alignment of (6,5) SWCNT and PFO−BPy. The density of states of (6,5) SWCNT with characteristic van Hove singularities of the valence ($v_1$, $v_2$, $v_3$) and conduction ($c_1$, $c_2$, $c_3$) band was based on ref 28 and shifted by the reported ionization potential.[16] The HOMO and LUMO energies of PFO−BPy indicated by red horizontal lines were reported by Jang et al.[29] The gray arrows are simplified representations for observed excitonic absorption bands $E_{11}$, $E_{22}$, and $E_{33}$. The inset shows the molecular structure of PFO−BPy.

## EXPERIMENTAL SECTION

**Surfactant-Based Carbon Nanotube Dispersion and Chirality Enrichment.** The preparation of aqueous (6,5) suspension is based on the pH-modulated aqueous two-phase extraction (ATPE) method.[41] Simply, a 20 mg portion of CoMoCAT SG65i SWCNTs powder (SouthWest Nanotechnologies, lot no. SG65i-L58) was suspended in 20 mL of aqueous 1% (m/v) DOC (BioChemica) by tip sonication (Weber Ultrasonics, 35 kHz, 16 W in continuous mode) for 1 h while immersed in an ice bath. The resulting dispersion was centrifuged at 45 560*g* (Beckman Optima L-80 XP, SW 40 Ti rotor) for 1 h, and the supernatant collected for ATPE. ATPE then was performed at a concentration of 4% (m/m) dextran (Mw 70 000 Da, TCI), 8% (m/m) PEG (Mw 6000 Da, Alfa Aesar) with 0.5% m/v SDS (Sigma-Aldrich), and 0.05% m/v DOC (Sigma-Aldrich). First, 16 μL of HCl (0.5 M) was added to a 16 mL ATPE system (with 0.8 mL of SWCNT suspension) to remove the nanotubes with a diameter larger than the (6,5). Next, a fresh mimic top phase was added along with some new HCl (20 μL) to obtain (6,5) on the top phase. Finally, the (6,5) enriched top phase was added to a fresh mimic bottom phase with 40 μL of sodium hypochlorite (NaClO, Honeywell) and 20 μL of compensated HCl to separate the metallic tubes with similar diameters. After centrifugation, the purified single chirality (6,5) remained in the bottom phase while the metallic tubes (7,4) partitioned to the top phase.

**Carbon Nanotube Dispersion by Polymer Wrapping.** As described previously,[2] nearly monochiral polymer-wrapped (6,5) SWCNTs were obtained by shear force mixing (Silverson L2/Air, 10 230 rpm, 72 h) of CoMoCAT raw material (Chasm Advanced Materials, SG65i-L58, 0.38 g $L^{-1}$) and poly-[(9,9-dioctylfluorenyl-2,7-diyl)-*alt*-(6,6′)-(2,2′-bipyridine)] (PFO−BPy, American Dye Source, Mw 40 000 Da, 0.5 g $L^{-1}$) in toluene. Aggregates were removed by centrifugation at 60 000*g* (Beckman Coulter Avanti J26XP centrifuge) for 2 × 45 min with intermediate supernatant extraction and final filtration through a poly(tetrafluoroethylene) (PTFE) syringe filter (5 μm pore size). To remove unbound PFO−BPy, the dispersion was passed through a PTFE membrane filter (Merck Millipore, JVWP, 0.1 μm pore size) to collect the SWCNTs and separate the unbound polymer. The SWCNT-coated membrane was further washed by immersion in toluene at 80 °C for 10 min. Finally, the washed, PFO−BPy-wrapped (6,5) SWCNTs (**Hybrid**) were redispersed from the membrane by bath sonication in a small volume of tetrahydrofuran (THF) for 30 min.

**Stationary and Transient Absorption Spectral Measurements.** Stationary UV−visible−NIR absorption spectra were measured on a V-770 (JASCO) spectrophotometer. Femtosecond transient absorption (TA) spectral measurements were performed on a commercial TA spectrograph (Helios Fire, Ultrafast Systems). The pump spectra centered at 1000, 576, or 350 nm were generated with a commercial optical parametric amplifier (TOPAS-Prime, Light Conversion), which was pumped by a regeneratively amplified femtosecond Ti:sapphire laser (Astrella, Coherent) centered at 800 nm, with a 4 kHz repetition rate, 78 fs pulse durations, and 1.6 mJ pulse energy. The spectra of pump pulses are shown in Figure S1 in the Supporting Information (SI). The spot size of the focused pump beam was about 250 μm. Typically, pump fluences were 200 μJ·$cm^{-2}$ for pulse energies of 100 nJ. The supercontinuum probe beam was generated in a sapphire substrate for the NIR detection (800−1350 nm) or in a calcium fluoride substrate for the UV−vis detection (330−650 nm). The pump beam was linearly polarized at the magic angle (54.7°) relative to the probe beam. TA spectra were corrected for the group velocity dispersion of the broad-band probe beam before analysis. All measurements were performed under ambient conditions.

**Spectroelectrochemical Measurement.** The spectroelectrochemical measurement was carried out on a Lambda 750 (PerkinElmer) UV−visible spectrophotometer combined with a CHI 660D (CH Instruments) potentiostat. The working electrode was an ITO glass (<10 Ω/square). The counter electrode was a platinum coil. The reference electrode was Ag/AgCl. The electrolyte is 0.1 M $n\mathrm{Bu_4NPF_6}$. Experiments were carried out at ambient temperature under the protection of nitrogen.

## ■ RESULTS

**Stationary Spectral Characterizations.** The stationary absorption spectra (Figure 2) of SWCNT/PFO−BPy **Hybrid**

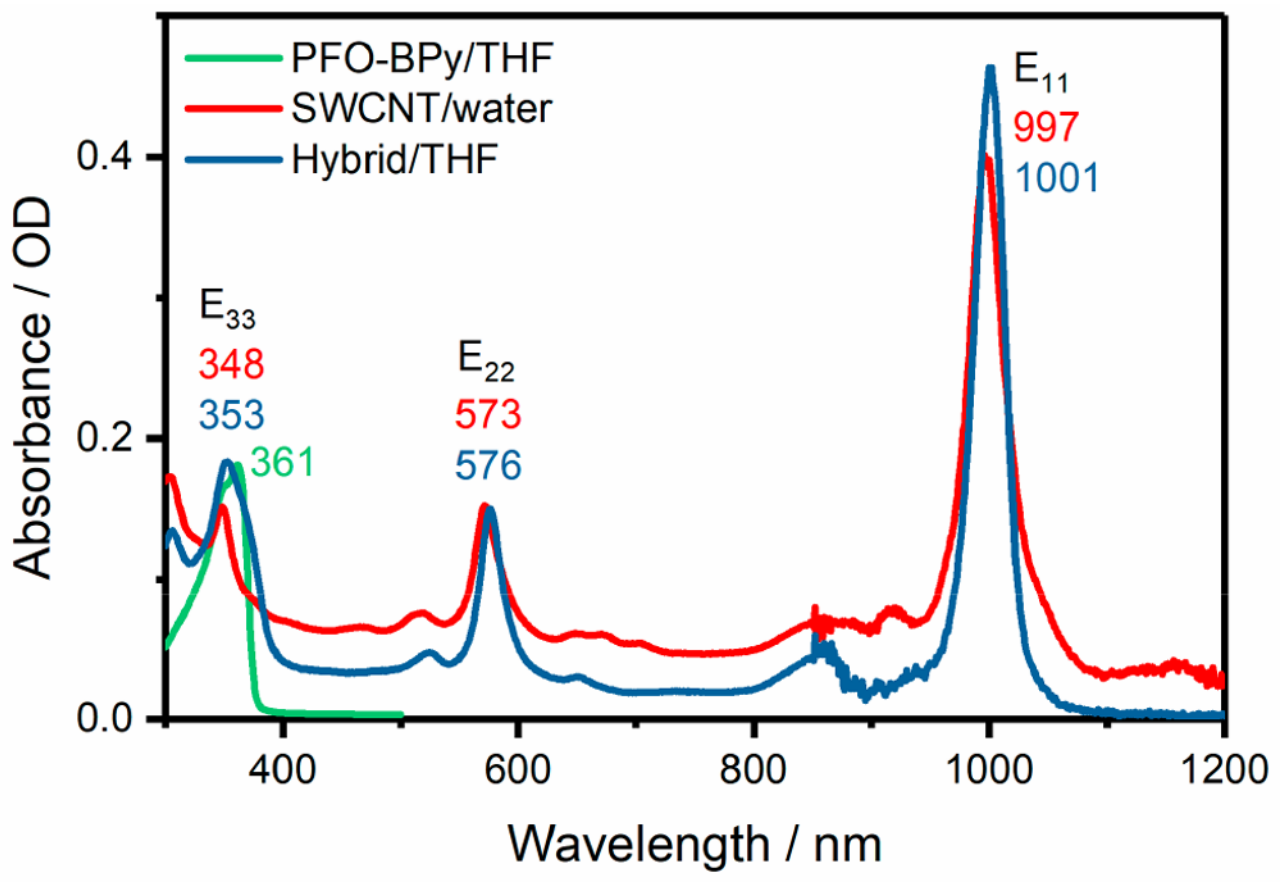

**Figure 2.** Stationary absorption spectra of surfactant-dispersed (6,5) **SWCNT** in water, PFO−BPy-wrapped (6,5) SWCNT **Hybrid** in THF, and PFO−BPy in THF. The positions of absorption peaks are marked with corresponding colors.

and surfactant-dispersed **SWCNT** show the typical excitonic transitions, i.e., $E_{00} \rightarrow E_{11}$, $E_{00} \rightarrow E_{22}$, and $E_{00} \rightarrow E_{33}$ ($E_{00}$ denotes the ground state in the exciton picture, as shown in Figure S1 in the Supporting Information), peaking around 1000, 576, and 350 nm, respectively. The absorption of PFO−BPy in the **Hybrid** system is observed as a shoulder on the red side of the $E_{00} \rightarrow E_{33}$ absorption band (see Supporting Information, Figure S2 for details). The visibility of the $E_{00} \rightarrow E_{33}$ absorption is due to the carefully reduced PFO−BPy concentration. Moreover, the chirality distributions of the **SWCNT** and **Hybrid** samples have been characterized by absorption spectroscopy (see Supporting Information, section B).[42]

**Transient Absorption Spectra of the SWCNT.** Figure 3 displays selected near-infrared (NIR) TA spectra of **SWCNT** upon the $E_{11}$, $E_{22}$, and $E_{33}$ resonant excitations. Due to the purity of the sorted (6,5) **SWCNT**, several known absorption features are clearly observed in the TA spectra. Upon $E_{11}$ excitation (Figure 3a) the dominant negative signal centered at ∼1000 nm arises from the $E_{00} \rightarrow E_{11}$ bleach, as reported previously.[43] The photoinduced absorption (PA) band centered at ∼1110 nm (∼1.12 eV) on the red side of the $E_{00} \rightarrow E_{11}$ bleach builds up within the instrumental response time and decays subsequently. This PA band has been assigned to the transition from exciton to biexciton ($E_{11} \rightarrow E_{11,BX}$).[39,44] Another PA band peaking at ∼1143 nm (∼1.08 eV) is evident after ∼10 ps and dominating the long-time scale spectra up to the limit of the measurement time window. Previous studies attributed this band to the triplet exciton absorption (${}^3E_{11} \rightarrow {}^3E_{nn}$).[45] Additionally, a broad PA band over 1200−1350 nm decays rapidly within the initial ∼1 ps. This spectral feature is commonly observed for the optical excitation into the $E_{11}$ excitonic band at high pump fluence and has been discussed to be related to multiple $E_{11}$ exciton interactions.[38,39,46,47] The PA band on the blue side of the $E_{00} \rightarrow E_{11}$ bleach is attributed to a transition from a dark $E_{11}$ state to the unbound two-exciton manifold.[48]

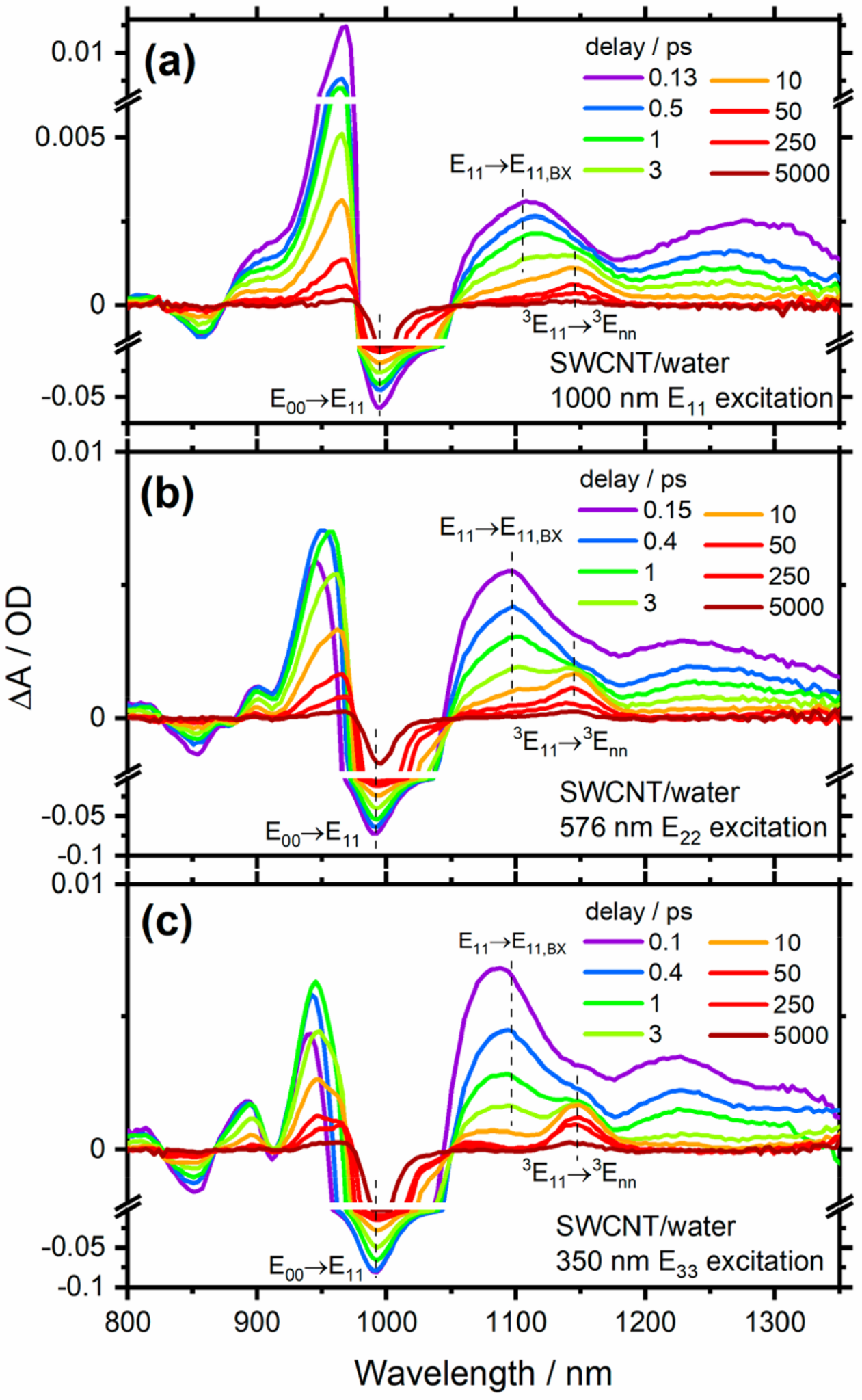

**Figure 3.** Selected TA spectra of **SWCNT** in water upon the (a) $E_{11}$, (b) $E_{22}$, and (c) $E_{33}$ excitations. Experimental conditions: (a) $\lambda_{ex}$ = 1000 nm, (b) $\lambda_{ex}$ = 576 nm, and (c) $\lambda_{ex}$ = 350 nm; pump energy: 100 nJ·pulse$^{-1}$. Dotted lines highlight major transition manifolds.

Resonant excitation into the $E_{22}$ or $E_{33}$ excitonic band of **SWCNT** results basically in the same spectral features (Figure 3b, c): the $E_{11} \rightarrow E_{11,BX}$ and ${}^3E_{11} \rightarrow {}^3E_{nn}$ transitions are still evident and show spectral line shapes and positions in agreement with those upon the $E_{11}$ excitation (vide supra). For excitons created by higher band transitions, such as $E_{22}$, the intersubband relaxation to the $E_{11}$ state has been shown to take place within 100 fs.[30,40,48,49] Upon $E_{22}$ or $E_{33}$ excitation, the $E_{00} \rightarrow E_{11}$ bleach with a large negative amplitude becomes broader and slightly blue-shifted, which results in larger overlap with the blue-side PA band. Similar spectral dynamics of **SWCNT** are observed in our results upon the excitation of either the $E_{11}$, $E_{22}$, or $E_{33}$ excitonic band.[44]

**Transient Absorption Spectra of the SWCNT/PFO−BPy Hybrid.** In the previous section, the intrinsic exciton dynamics of (6,5) SWCNTs upon different excitonic−transition excitations have been shown. Our focus now turns to the excited-state dynamics of the (6,5) SWCNT/PFO−BPy formed heterojunction to find evidence for PCT. Selected NIR TA spectra of the **Hybrid** upon the $E_{11}$, $E_{22}$, and $E_{33}$ excitations are shown in Figure 4. Besides the known $E_{11} \rightarrow E_{11,BX}$ (~1100 nm, ~1.13 eV) and $^3E_{11} \rightarrow {}^3E_{nn}$ (~1160 nm, ~1.07 eV) transitions, an additional broad PA band emerges on the red side of the $E_{00} \rightarrow E_{11}$ bleach in the **Hybrid** upon $E_{22}$ and $E_{33}$ excitation (highlighted by shaded areas in Figure 4b, c). It covers the spectral range of ~1050−1200 nm that overlaps with the absorption of $E_{11} \rightarrow E_{11,BX}$ and $^3E_{11} \rightarrow {}^3E_{nn}$ transitions and lasts for tens of picoseconds. Furthermore, on this time scale, we note that the $E_{00} \rightarrow E_{11}$ bleach in the TA spectra of the **Hybrid** upon the $E_{33}$ excitation manifests a dynamic blue-shift of up to ~5 nm. We summarize the peak dynamics in Figure 5. The time-dependent peak-shifting of the $E_{00} \rightarrow E_{11}$ bleach was extracted from the TA spectra by Gaussian-peak fitting in the energy domain within the spectral region of the dominant bleach band centered at ~1000 nm. Considering that the exciton density is strongly dependent on the excitation fluence (section I), we investigated this peak shift for the **SWCNT** and **Hybrid** with various excitation fluences (see Supporting Information, section C for fluence-dependent TA spectra).

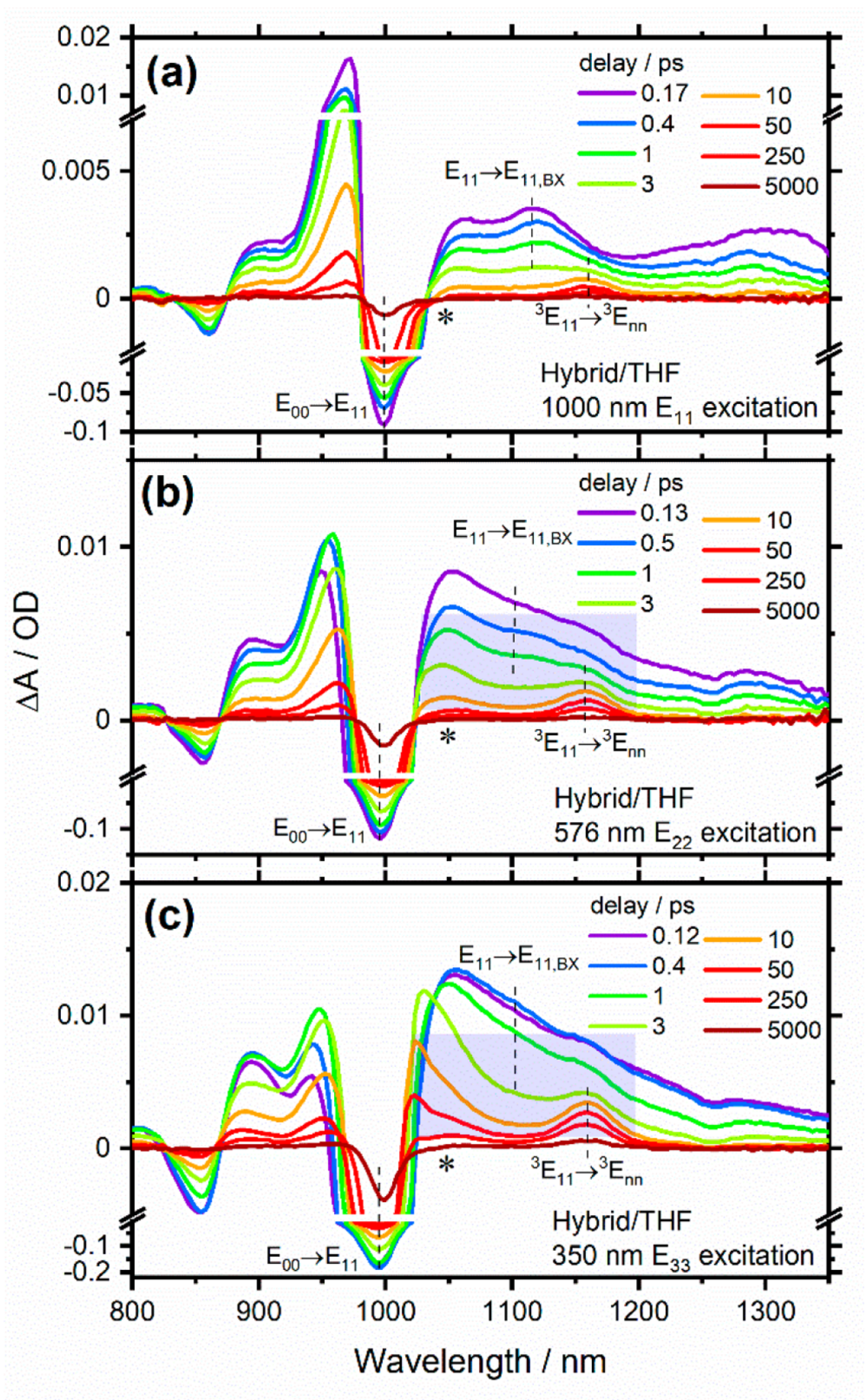

**Figure 4.** Selected TA spectra for the **Hybrid** in THF upon the (a) $E_{11}$, (b) $E_{22}$, and (c) $E_{33}$ excitations. Experimental conditions: (a) $\lambda_{ex}$ = 1000 nm, (b) $\lambda_{ex}$ = 576 nm, and (c) $\lambda_{ex}$ = 350 nm; pump energy: 100 nJ·pulse$^{-1}$. Dotted lines highlight major transition manifolds. The shaded shapes indicate the absorption signature of the suspected SWCNT polaron. The asterisks (*) denote the wavelength of 1050 nm.

As shown in Figure 5a, the peak position of the $E_{00} \rightarrow E_{11}$ bleach for the **SWCNT** upon the $E_{11}$ excitation remains stable across the investigated time window. For the **SWCNT** upon $E_{22}$ or $E_{33}$ excitation (Figure 5b and 5c, respectively), the bleach peak shows a slight and smooth red-shift through the entire time window. This trend is widely observed when tracking the dynamics of excitonic band bleach recovery in SWCNTs, and it can be explained by intersubband and intrasubband relaxation.[39,50] Upon exciting the **Hybrid** with different pump wavelengths, the initial peak position at ~0.1 ps shifts slightly, which matches well with the trend observed in the **SWCNT**. However, the peak-shifting dynamics are significantly different for the **Hybrid**. Besides the smooth red-shift through the entire time window, a dynamic blue-shift of the $E_{00} \rightarrow E_{11}$ bleach is particularly evident within the time delay from 1 to 50 ps when the **Hybrid** is pumped at $E_{33}$ (Figure 5f). Notably, the extent of the dynamic blue-shift progressively becomes larger with increasing excitation fluence. When the peak-shifting curves of the **Hybrid** upon $E_{22}$ excitation (Figure 5e) are examined, although these curves do not show an obvious blue-shift as $E_{33}$ excitation, the peak position plateaus on the same time scale at the highest excitation energy. For the **Hybrid** upon $E_{11}$ excitation, we hardly observe this trend (Figure 5d), and it behaves almost the same as the **SWCNT** upon $E_{11}$ excitation. The additional peak dynamics of the $E_{00} \rightarrow E_{11}$ bleach in the **Hybrid** upon $E_{33}$ (and less pronounced upon $E_{22}$) excitation indicate an additional quasiparticle with a blue-shifted $E_{00} \rightarrow E_{11}$ bleach and an observable buildup time, which may be the PCT product. One point to note is that the dynamic blue-shift in tens of picoseconds does not directly reflect real population dynamics. It only suggests that the share of this new species is increasing among all quasiparticles, and the population of each transient species may be decaying individually on that time scale.

**Absorption Features of Charge Transfer Products.** To identify the PCT in the **Hybrid**, we examine the spectroscopic features of potential PCT transient products.[22,51] Charged transient products in SWCNTs have been demonstrated to be two kinds of quasiparticles: (a) the polaron, which describes a conduction electron (or hole) strongly coupled with the lattice ions,[52] or (b) the trion, which is a three-body charge-exciton bound state.[38,47,53,54]

The (6,5) SWCNT polaron can be created in steady state by two methods, namely redox-chemical doping[22,25,55−57] and electrochemical doping.[6,58,59] We carried out the redox-chemical hole-doping of the **Hybrid** dispersion with $NOBF_4$ (+1.00 V vs Fc/Fc$^+$ in $CH_2Cl_2$)[60] as a one-electron oxidant.[61] As shown in Figure 6a, along with the increasing doping level, the stationary absorption of the $E_{00} \rightarrow E_{11}$ transition (~1000 nm) decreases dramatically and exhibits a blue-shift in wavelength. On the red side of the dominant $E_{00} \rightarrow E_{11}$ transition, stationary spectra of the doped **Hybrid** feature a broad absorptive band extending from 1030 to 1200 nm, including a peak around ~1150 nm. The weak absorption at 1150 nm before the oxidative titration ([$NOBF_4$] = 0 $\mu$M) results from the slight p-doping of SWCNTs in the air.[62] We expect that the polymer in the **Hybrid** remains unoxidized during the titration, since the HOMO energy of PFO−BPy is stabilized by 1.4 eV with respect to the first valence band of (6,5) SWCNT. The previously reported electrochemical doping for the **Hybrid** in films shows a very similar trend to the redox doping.[6] Note that for very high doping levels, the entire NIR absorption of the **Hybrid** is strongly bleached. We conclude that the above spectral features describe the (6,5) SWCNT hole-polaron absorption transition, denoted as $E^{+}_{00} \rightarrow E^{+}_{11}$.

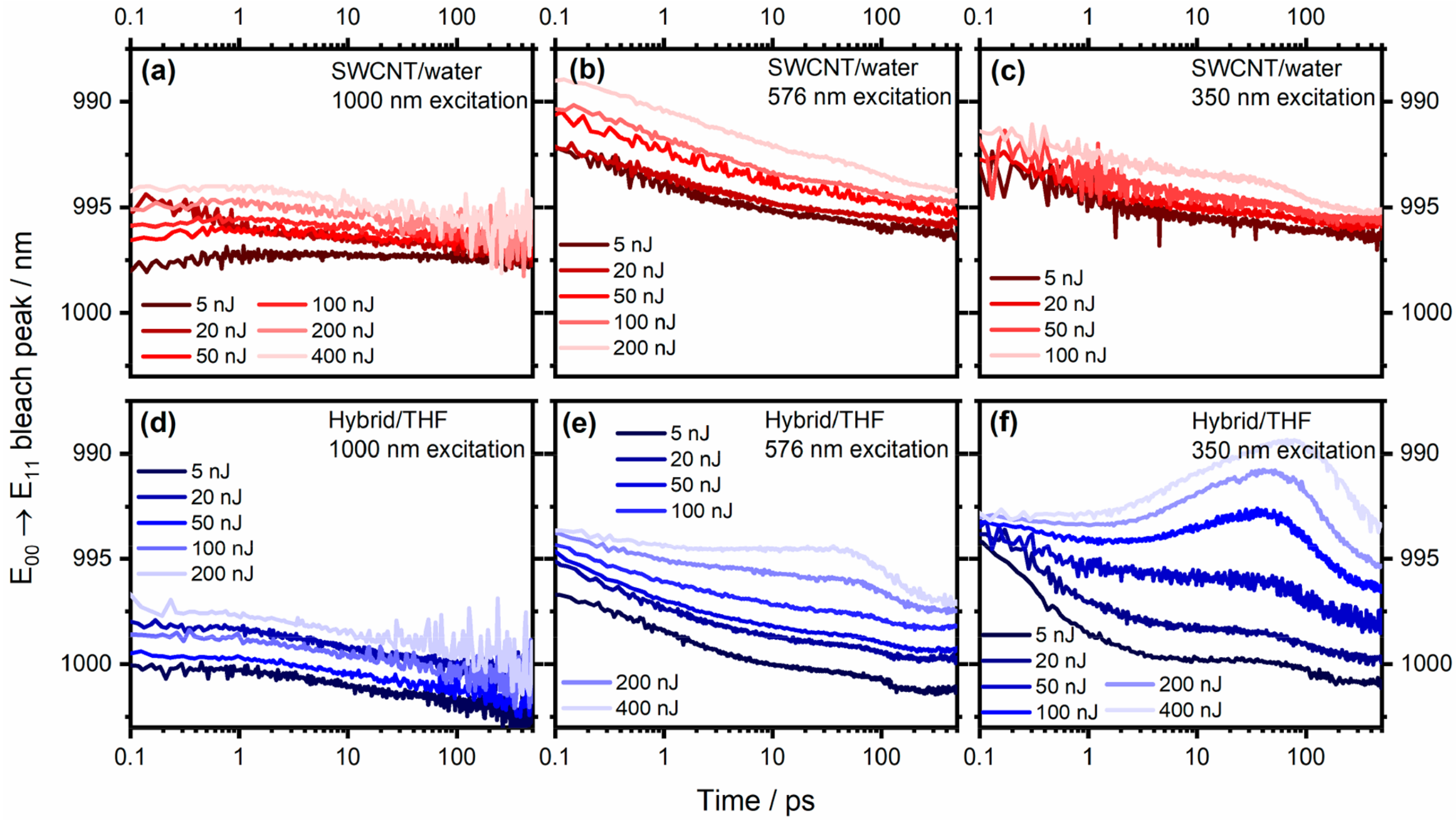

**Figure 5.** Pump-energy-dependent peak-shifting dynamics of the $E_{00} \rightarrow E_{11}$ bleaching in TA spectra of the **SWCNT** (a, b, c) and the **Hybrid** (d, e, f) in the time window of 0.1−500 ps. Excitation wavelength and corresponding pump energy per pulse are given in legends. Due to dispersion instability under high pump fluences, TA spectra of **SWCNT** are unavailable at higher fluences in b and c.

In order to examine the spectral signature of trions, we conducted the TA measurement on the heavily hole-doped ($[NOBF_4]$ ∼ 128 $\mu$M) **Hybrid** under $E_{11}$ excitation.[47] As shown in Figure 6b, the initial TA spectra show the $E_{00} \rightarrow E_{11}$ bleach at ∼1000 nm and $E^{+}_{00} \rightarrow E^{+}_{11}$ bleach at ∼1150 nm. The $E^{+}_{00} \rightarrow E^{+}_{11}$ bleach decreases within ∼1 ps, and meanwhile a new absorptive species centered at ∼1190 nm (∼1.04 eV) is formed. This band has been assigned to a positive trion ($Tr^{+}_{11} \rightarrow Tr^{+}_{nn}$) absorption of (6,5) SWCNTs.[38,47] The spectral signature of trions in the **Hybrid** is consistent with that observed in (6,5) SWCNT superstructures reported by Therien et al.[38,47]

## ■ DISCUSSION

**Observation of Charge Transfer in SWCNT/PFO−BPy Hybrid.** Combining the results of redox-chemical and electrochemical doping (vide supra), we can summarize the spectroscopic features of the (6,5) SWCNT hole-polaron ($E^{+}_{00} \rightarrow E^{+}_{11}$) in comparison with the excitonic absorption spectrum of neutral (6,5) SWCNTs. The (6,5) SWCNTs hole-polaron features a bleached and blue-shifted $E_{00} \rightarrow E_{11}$ transition and an additional absorption band covering 1030−1200 nm, the line shape of which strongly depends on the charge carrier level.[25,55,57−59,63] The electronic absorption transitions of oxidized nanotubes are explained by the electron depletion of the top of the valence band, which results in an increase in the $E_{00} \rightarrow E_{11}$ transition energy and leads to additional electronic transitions.[25,59] It is noteworthy that studies by electron paramagnetic resonance (EPR) spectroscopy reveal that the unpaired electrons in lightly reduced SWCNTs are relatively free and fast-relaxing.[20] This is certainly valid for metallic SWCNTs but has been corroborated as well for semiconducting SWCNTs.[64] Thus, the "polaron" in SWCNTs is relatively delocalized, which may prolong the lifetime of the charge-separation state.

As shown in Figure 5, TA spectra of the **Hybrid** feature a dynamic blue-shift of the $E_{00} \rightarrow E_{11}$ bleach in picoseconds following the $E_{22}$ and $E_{33}$ excitations, which strongly suggests the formation of the SWCNT polaron ($E^{\pm}_{00} \rightarrow E^{\pm}_{11}$). Although the triplet absorption also contributes to the absorption around 1160 nm where an absorption peak of $E^{\pm}_{00} \rightarrow E^{\pm}_{11}$ is located, the TA spectra of **SWCNT** indicate that the formation of triplet excitons is independent of the excitation energy and fluence. According to the reported time constant of intersystem crossing of ∼20 ps in (6,5) SWCNTs,[45] the broad absorption band at 1030−1200 nm and the peak around 1160 nm formed in the first few picoseconds in the **Hybrid** upon high-energy excitation should be attributed to the $E^{\pm}_{00} \rightarrow E^{\pm}_{11}$ transition. However, the spectral overlap between $E^{\pm}_{00} \rightarrow E^{\pm}_{11}$, $E_{11} \rightarrow E_{11,BX}$, and $^{3}E_{11} \rightarrow {}^{3}E_{nn}$ transitions (compare Figure 4 with Figure 6a) impedes a spectral disentanglement of the SWCNT polaron dynamics via global analysis of the TA dynamics of the **Hybrid**. Thus, we select the TA kinetics at 1050 nm, where an isosbestic point with almost zero $\Delta A$ is located in the TA spectra of the **SWCNT**, to further investigate the formation of the SWCNT polaron in the **Hybrid**. As shown in Figure 7a−c, the normalized kinetic traces for the **Hybrid** pumped at very low pump energy (5 or 20 nJ·pulse$^{-1}$) show a smooth decay which is generally the same for the $E_{11}$, $E_{22}$, and $E_{33}$ excitations. This decay component arises from the absorption edge of $E_{11} \rightarrow E_{11,BX}$ (centered at 1100 nm) transition. When the pump energy is increased from 50 to 400 nJ·pulse$^{-1}$, the dynamics of the SWCNT polaron emerge. It becomes progressively evident in the **Hybrid** upon $E_{22}$ and $E_{33}$ excitation (Figure 7b, c) and overlaps with the $E_{11} \rightarrow E_{11,BX}$ absorption. The absorption amplitude reaches its maximum at ~3 ps and slowly decays over a few hundred picoseconds. The spectroscopic features of SWCNT polaron absorption over 1050−1200 nm can be observed in the normalized TA spectra of the **Hybrid** at the probe delay of 3 ps (Figure 7e, f), matching well with the absorptive feature of the SWCNT hole-polaron ($E^+_{00} \rightarrow E^+_{11}$, shown in blue lines) obtained by chemical doping (vide supra). In the TA spectra of the **Hybrid** upon $E_{11}$ excitation, the signature of the SWCNT polaron is hardly observed in the respective kinetics and spectra (Figure 7a, d). In addition, using the spectroelectrochemical method, the PFO−BPy polaron was prepared and observed in the steady-state spectrum. As one would expect, the formation of the PFO−BPy polaron is also observed in the UV region of the **Hybrid** TA spectra around 380 nm (see Supporting Information, sections E and F). Hence, we conclude that a PCT reaction takes place in the SWCNT/PFO−BPy **Hybrid** system, forming an interfacial charge-separated state.[22]

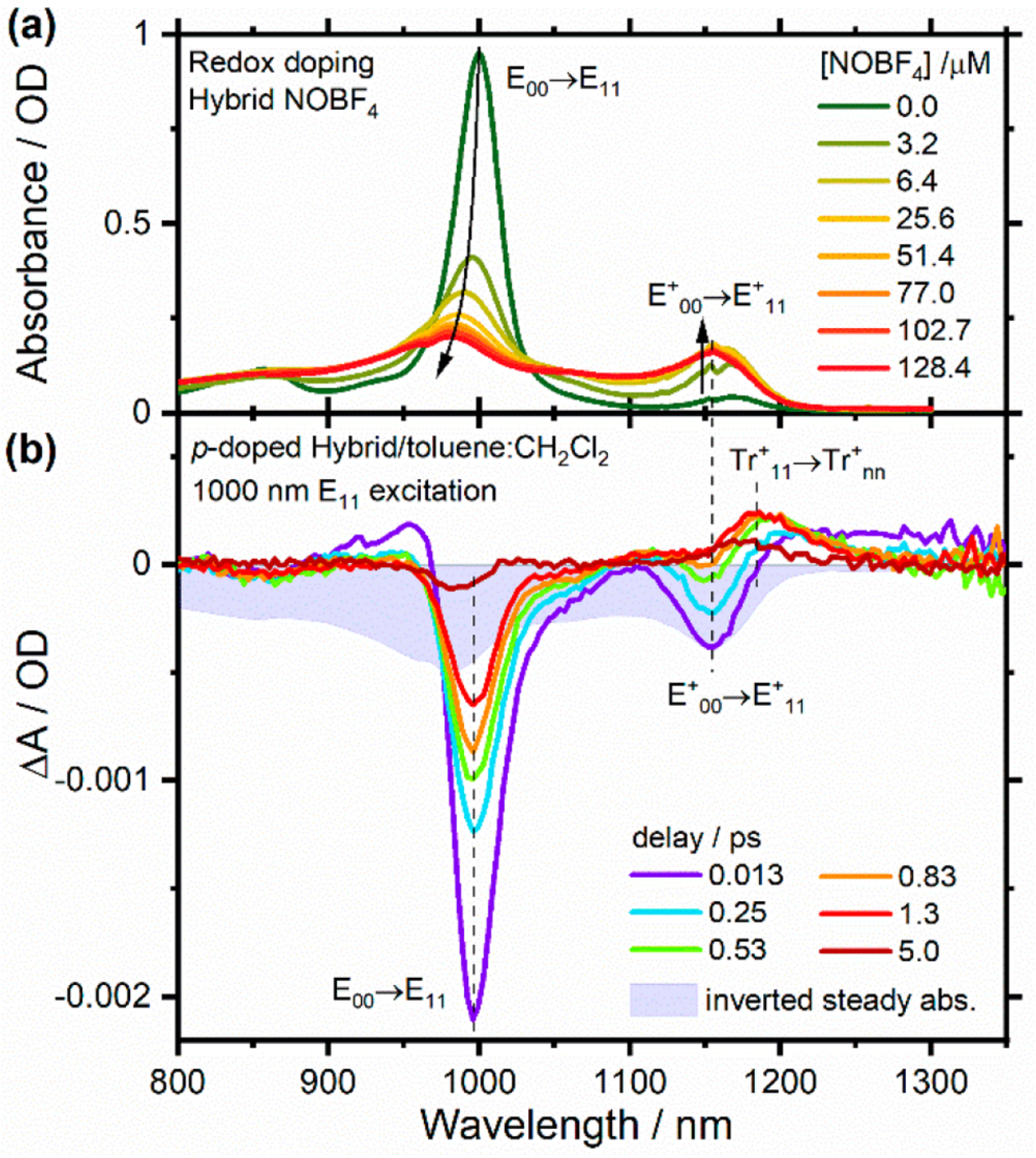

**Figure 6.** (a) NIR stationary absorption spectra monitor the oxidative titration of the **Hybrid** with $NOBF_4$ in toluene:$CH_2Cl_2$ (ratio 1:1) mixed solution. Experimental conditions: [(6,5) SWCNT] ~ 2.74 nM; SWCNT length ~1000 nm; optical path length = 10 mm. (b) Selected TA spectra for a heavily hole-doped ([$NOBF_4$] ~ 128 $\mu$M) **Hybrid** in toluene:$CH_2Cl_2$ (ratio 1:1) mixed solution. Experimental conditions: $\lambda_{ex}$ = 1000 nm, i.e., in resonance with $E_{11}$; pump energy = 50 nJ·pulse$^{-1}$. Scaled steady-state absorption spectrum (inverted shaded shape) is shown for comparison.

It is worth adding that we have not observed any spectral evidence of trion formation, as previously identified from the hole-doped **Hybrid** solution (Figure 6b), in the TA spectra of undoped **SWCNT** or **Hybrid**.

In the **Hybrid**, the direction of PCT can be deduced by analyzing the excitation distribution between the two components. In the case of $E_{11}$ or $E_{22}$ excitation of the **Hybrid**, the pump pulse centered at 1000 or 576 nm is far away from the UV resonant absorption of PFO−BPy. Thus, the PCT in both cases purely originates from the excited (6,5) SWCNT to the PFO−BPy. In the case of $E_{33}$ excitation, the pump spectrum centered at 350 nm is resonant with (6,5) SWCNT as well as with PFO−BPy. Since PCT from excited PFO−BPy to the (6,5) SWCNTs is also energetically favorable in this case, one needs to evaluate the excitation contribution from (6,5) SWCNT and PFO−BPy in the **Hybrid** by comparing the initial UV−vis TA spectra of the **Hybrid**, **SWCNT**, and PFO−BPy excited at 350 nm. As shown in Figure 8, the TA spectrum of PFO−BPy at 0.2 ps clearly manifests a negative bleach peak (~365 nm) and a negative stimulated emission peak (~390 nm) (see Supporting Information, section D for the TA spectral analysis of PFO−BPy). However, the above TA spectral features from excited PFO−BPy are hardly observed in the TA spectrum of the **Hybrid** at the same time delay (see Supporting Information, section E for a comparison of the full dynamics). The **Hybrid** TA spectrum only retains the excitation features of the **SWCNT**, especially for the region of 350−450 nm, and the excitation contribution of PFO−BPy is minimal in comparison to the excitation of (6,5) SWCNTs in the **Hybrid**. Thus, we conclude that although the PCT from excited PFO−BPy to (6,5) SWCNT cannot be completely ruled out, the current analysis shows that, in the case of $E_{33}$ (350 nm) excitation of the **Hybrid**, the PCT from excited (6,5) SWCNT to PFO−BPy is absolutely dominant.

An additional point to address is whether an electron or hole transfer takes place. Since the electron and hole in SWCNT have a similar effective mass, electron- and hole-polaron will show very similar absorptive signatures at the same doping level.[65] However, the energy level alignment in the **Hybrid** (Figure 1) implies that the electron transfer from the third conduction subband ($c_3$) of (6,5) SWCNT to the LUMO of PFO−BPy has a larger driving force than the hole transfer from the third valence subband ($v_3$) to the HOMO. Therefore, photoinduced electron transfer from (6,5) SWCNT to PFO−BPy is more likely to take place than hole transfer.

**Charge Transfer Assisted by Auger Recombination.** As discussed above, when the wide-bandgap PFO−BPy acts as a charge acceptor in the **Hybrid**, the PCT from the $E_{11}$/$E_{22}$-excited (6,5) SWCNT to PFO−BPy is energetically unfavorable (Figure 1). However, when multiple excitons are present on a SWCNT in a high-excitation-density regime, strong exciton−exciton interactions lead to Auger recombination of excitons, also known as exciton−exciton annihilation (EEA), in which one exciton recombines to the ground state and the other is promoted to a higher excitonic state or dissociates into free charge carriers.[34,35,39,66] Hence, Auger recombination of excitons can be regarded as a potential energy up-conversion process that forms one high-energy exciton through absorbing two low-energy photons. Under this assumption, the occurrence of PCT driven by $E_{22}$ (576 nm, 2.15 eV) excitation in the **Hybrid** indicates that $E_{33}$ (350 nm, 3.54 eV) or higher-energy excitons are generated via Auger recombination of excitons, thus making the reaction energetically favorable. For the **Hybrid** upon $E_{11}$ (1000 nm, 1.24 eV) excitation, however, the annihilation of two $E_{11}$ excitons is energetically insufficient to directly promote an $E_{33}$ exciton (350 nm, 3.54 eV). Consequently, the PCT from the $E_{11}$-excited (6,5) SWCNT to PFO−BPy is quite inefficient as observed in Figure 7. According to previous reports on fluence-dependent dynamics in SWCNTs,[67] we conclude from the estimated exciton density (see Supporting Information, section I) that the exciton−exciton Auger process is unlikely to occur under $E_{22}$ excitation with less than 20 nJ·pulse$^{-1}$. As shown in Figure 7, the PCT characteristics are hardly observable under these conditions.

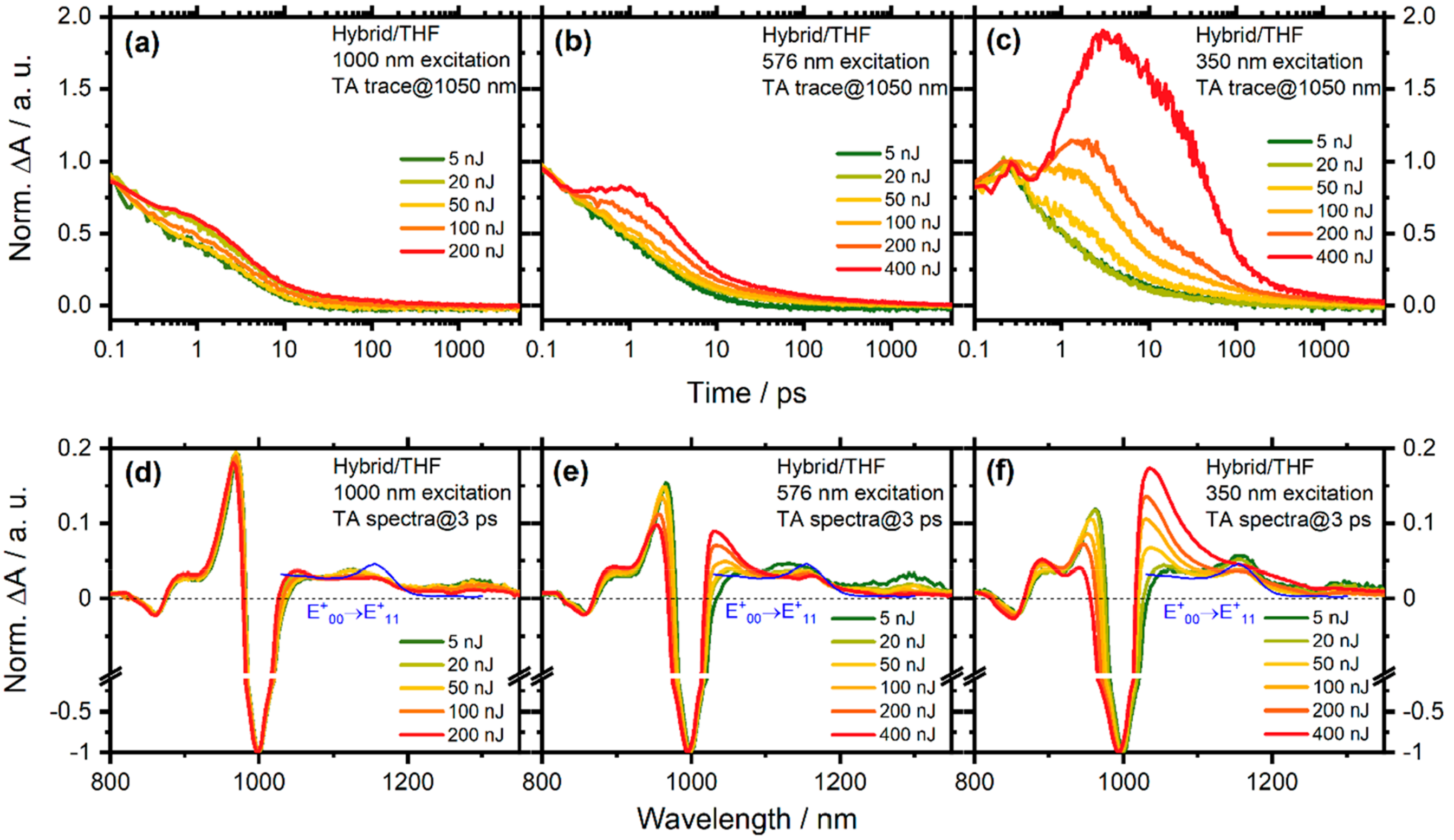

**Figure 7.** Normalized pump-energy-dependent TA traces at 1050 nm for the **Hybrid** in THF upon (a) $E_{11}$, (b) $E_{22}$, and (c) $E_{33}$ excitation. Note that traces were normalized by the $\Delta A$ amplitude at 0.1−0.2 ps considering the instrumental response. Normalized pump-energy-dependent TA spectra for the **Hybrid** in THF at a time delay of ~3 ps upon the (d) $E_{11}$, (e) $E_{22}$, and (f) $E_{33}$ excitations. Note that spectra were normalized at the $E_{00} \rightarrow E_{11}$ bleaching maximum. The blue lines represent the stationary absorption feature of the (6,5) SWCNT hole-polaron obtained by redox-chemical doping ($[NOBF_4]$ ~ 128.4 $\mu$M, shown in Figure 6a).

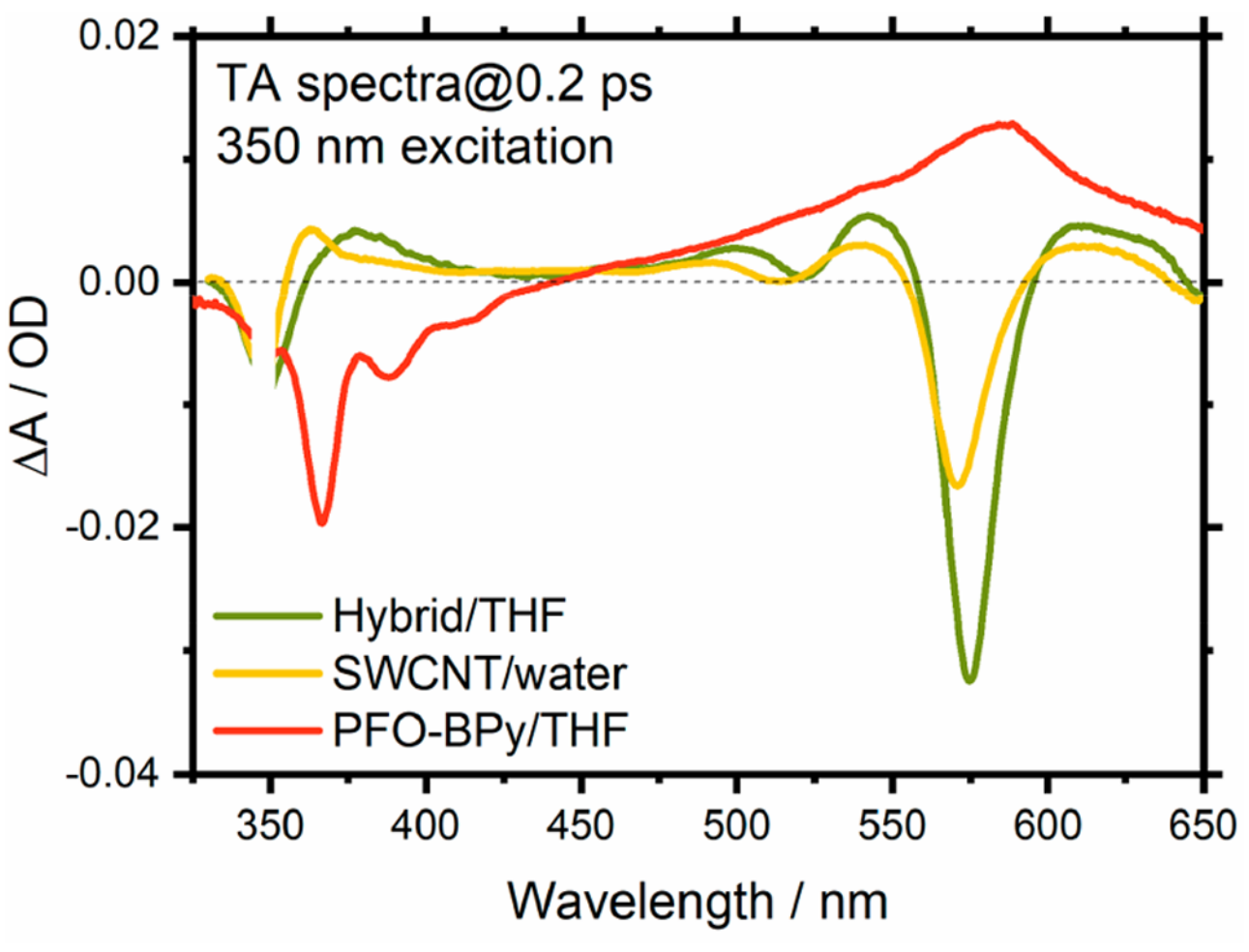

**Figure 8.** UV−vis TA spectra for the **Hybrid** in THF, **SWCNT** in water, and PFO−BPy in THF at a time delay of 0.2 ps. Excitation wavelength: 350 nm. Pump energy: 100 nJ·pulse$^{-1}$.

Due to the effect of exciton dissociation in SWCNTs, the elementary excitation specifically involved in the Auger recombination may not be limited to excitons but can also involve unbound charge carriers, i.e., electrons and holes (e, h). The up-converted product in the Auger recombination of charge carriers is a high-energy carrier (electron or hole), while the product in the Auger recombination of excitons is a high-energy exciton (schematically shown in Figure 9a). It has been shown that these two mechanisms can be distinguished by their distinct population kinetics.[34] Auger recombination of charge carriers, as a three-particle process, can be described by a rate equation as

$$\frac{dn_{e,h}(t)}{dt} = -\frac{1}{3}\gamma_A n_{e,h}^3(t) \quad (1)$$

where $n_{e,h}(t)$ is the population of charge carriers, and $\gamma_A$ is the rate constant of the Auger recombination of charge carriers. The solution of eq 1, $[n_{e,h}(0)/n_{e,h}(t)]^2 - 1 = 2/3\gamma_A n_{e,h}^2(0)t$, where $n_{e,h}(0)$ denotes the initial population of charge carriers, predicts a linear dependence between the reciprocal of the charge carrier population squared and the delay time. Auger recombination of excitons, as a two-particle process, can be described as

$$\frac{dn_{ex}(t)}{dt} = -\frac{1}{2}\gamma_{EEA} n_{ex}^2(t) \quad (2)$$

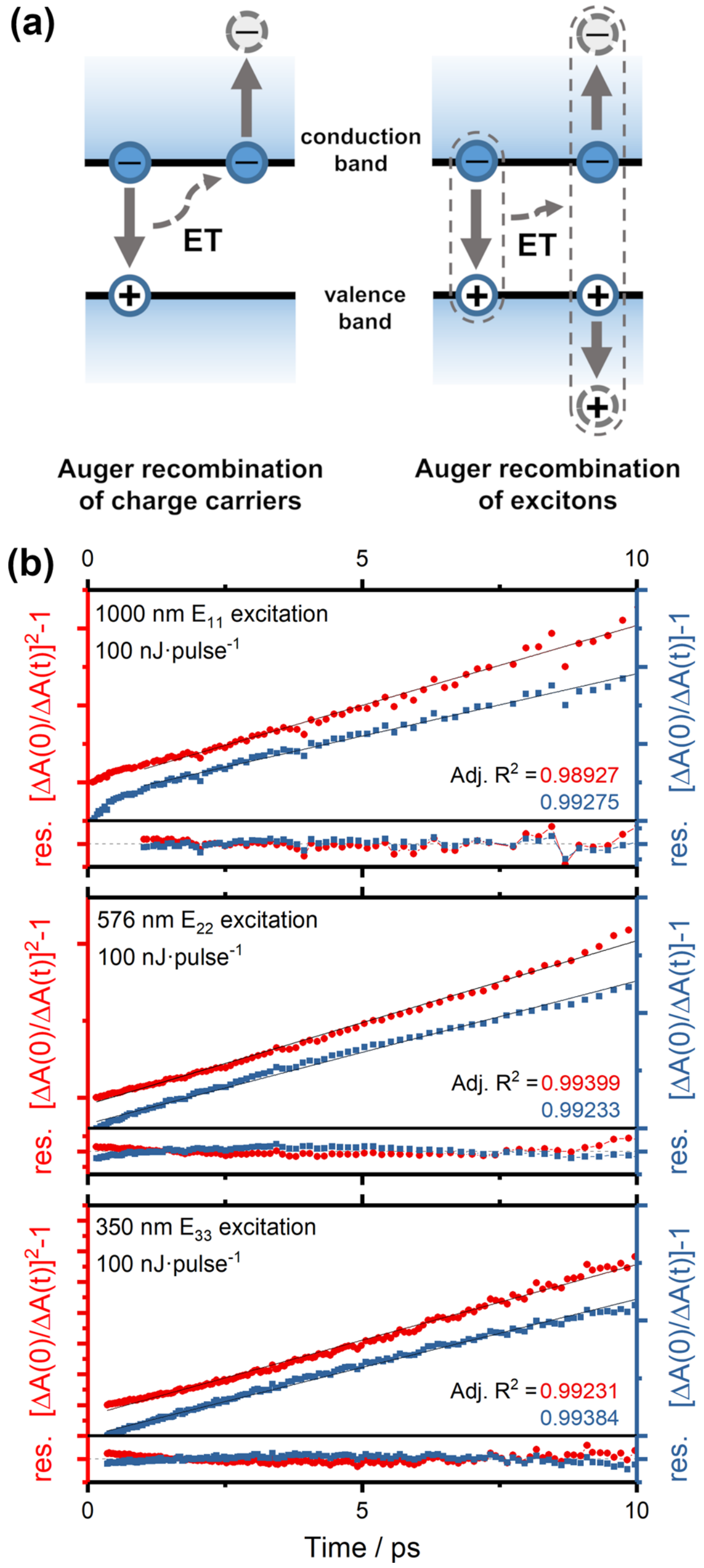

**Figure 9.** (a) Schematic description of the Auger recombination of charge carriers (left) and excitons (right). ET denotes energy transfer. (b) Kinetics of the integral $E_{00} \rightarrow E_{11}$ bleaching in the TA spectra of the **Hybrid** in THF upon the $E_{11}$, $E_{22}$, and $E_{33}$ excitations, plotted as $\{[\Delta A(0)/\Delta A(t)]^2 - 1\}$ (red dots, left axis) and $\{[\Delta A(0)/\Delta A(t)] - 1\}$ (blue squares, right axis). Traces are shifted by different offset on the vertical axis for a better comparison. Solid black lines represent the results of the linear fitting. Adjusted $R$-squared (Adj. $R^2$) and fitting residuals are shown with corresponding colors. Note that the fitting of the $E_{11}$-excited TA trace started from 1 ps to exclude the nondiffusion-controlled rapid annihilation.[35] Pump energy: 100 nJ·pulse$^{-1}$.

where $n_{ex}(t)$ is the population of excitons, and $\gamma_{EEA}$ is the rate constant of the Auger recombination of excitons. The solution of eq 2, $[n_{ex}(0)/n_{ex}(t)] - 1 = 1/2\gamma_{EEA}n_{ex}(0)t$, where $n_{ex}(0)$ denotes the initial population of excitons, predicts a linear dependence between the reciprocal of the exciton population and the delay time.

To analyze the time dependence of the bleach signal in the initial 10 ps of **Hybrid** upon $E_{11}$, $E_{22}$, and $E_{33}$ excitations, the $E_{00} \rightarrow E_{11}$ bleaching band in the TA signal is integrated, and its kinetics are plotted as shown in Figure 9b (pump energies of 100 nJ·pulse$^{-1}$) and section G (pump energies of 2−400 nJ·pulse$^{-1}$) in the Supporting Information. The linear fitting of $[\Delta A(0)/\Delta A(t)]$ and $[\Delta A(0)/\Delta A(t)]^2$ shows, however, comparable adjusted $R$-squared and fitting residuals for all excitation photon energies and excitation fluences. This result suggests that Auger recombination in the **Hybrid** in the initial 10 ps is not purely excitonic, but the Auger recombination of charge carriers coexists as well.

It is worth noting that when the pump photon energy increases from 1.24 eV (1000 nm, $E_{11}$ excitation) to 2.15 eV (576 nm, $E_{22}$ excitation) and 3.54 eV (350 nm, $E_{33}$ excitation), the charge carrier nature does not obviously replace the excitonic nature as dominant in the elementary excitations of the **Hybrid** in THF. Besides, the trion, as an indication of efficient free carrier generation (FCG), is absent as well in the TA spectra of the undoped **Hybrid** as discussed above. All these results indicate that the FCG of the **Hybrid** in THF is relatively inefficient. The exciton binding energy in carbon nanotubes increases with decreasing solvent dielectric permittivity,[68] which entails that FCG is highly sensitive to the electrostatic environment. Efficient FCG in pristine SWCNTs is usually observed in high dielectric permittivity microenvironments, such as polar solvents, ionic surfactants, or ionic semiconducting polymers.[33,38,69,70] Therien et al. pointed out the adverse effect of the low dielectric environment on FCG in SWCNTs by quantitatively analyzing the FCG efficiency in mixtures of $D_2O$ ($\varepsilon$ = 78.5) and MeOH ($\varepsilon$ = 32.6).[38] Therefore, we deduce that the FCG in the **Hybrid** may be suppressed by the low dielectric solvent, THF ($\varepsilon$ = 7.5). The wrapping PFO−BPy, as a weakly polar polymer, also provides a low dielectric microenvironment for SWCNTs, which is unfavorable to efficient FCG as well. The influence of environmental effects on all-optical FCG in polymer-wrapped SWCNTs still needs further study.

Beside Auger recombination, high-energy and high-fluence excitations may also lead to other nonlinear effects in SWCNTs. The multiple-exciton generation (MEG), as the opposite process of the Auger recombination, forms two excitons by absorbing one photon with energy higher than twice the bandgap.[36] MEG is, however, an intrinsic feature of SWCNTs and should thus be observable in both surfactant-dispersed and polymer-wrapped SWCNTs. The lack of any obvious changes in the line shape of the spectra caused by MEG upon $E_{33}$ excitation at high pump fluence for pure SWCNTs shows that MEG is not present. Furthermore, MEG as an energy down-conversion process should not contribute to the PCT from excited SWCNTs to wide-bandgap PFO−BPy. Finally, while two-photon absorption (TPA) is an energy up-conversion process which may favor the PCT, TPA coefficients in pristine semiconducting SWCNTs have been reported to be very low.[71]

**Charge Separation and Recombination Dynamics.** Finally, reaction time constants for photoexcited interfacial charge separation ($\tau_{CS}$) and thermal charge recombination ($\tau_{CR}$) in the SWCNT/PFO−BPy **Hybrid** were estimated by fitting the fluence-dependent kinetic traces (see Supporting Informaton, section H). The buildup of the SWCNT polaron signature takes place in $\tau_{CS} \approx 0.9$ ps, which is consistent with the time scale of reported electron transfer in perylenediimide-based polymer-wrapped carbon nanotube superstructures ($\tau_{CS} \approx 0.4$ ps).[22] The decay of the SWCNT polaron manifests in a multiphase process, which was reproduced by three-exponential kinetics ($\tau_{CR} \approx 3$, 50, and 1000 ps). However, the UV−vis TA spectra of the **Hybrid** corroborate that the decay of the PFO−BPy polaron is on a time scale of a few picoseconds and does not feature a nanosecond-lived component (see Supporting Information, section E). The charge recombination path should not be limited to the direct recombination between the SWCNT hole polaron and PFO−BPy electron polaron. The Type-I heterojunction can efficiently funnel the electron on the LUMO of PFO−BPy into the conduction band of the SWCNT, which is independent of whether the SWCNT is charged or neutral. Consequently, the PFO−BPy polaron features a significantly shorter decay time compared with that of the SWCNT polaron. When electrons back-transfer to neutral SWCNTs, SWCNT electron polarons will be formed and coexist with the hole polarons because of the high migration rate and large delocalization length of the SWCNT polaron.[25,72] Therefore, we deduce that the fast decay of the SWCNT polaron on a time scale of ∼3 ps might originate from the direct charge recombination between the closely associated SWCNT electron polaron and PFO−BPy hole polaron. The delocalized SWCNT electron and hole polarons, which migrate along the nanotubes backbone, prolong the final charge recombination to the subnanosecond time scale.[22,73−75]

Through comparison of the spectra of the SWCNT polaron produced by chemical oxidation and the maximal $E_{00} \rightarrow E_{11}$ blue-shift in the TA spectra (see Supporting Information, section J), we estimate that PCT induces a doping level higher than the equivalent of ∼8 $\mu$M $NOBF_4$ in the SWCNTs. However, because of the spectral overlap in the TA spectra, especially at the delay time when the population of the SWCNT polaron reaches its maximum at around 1−3 ps, we could not extract the pure $E_{00} \rightarrow E_{11}$ bleach of the SWCNT polaron. Thus, the degree of charge transfer in the PCT reaction cannot be quantified precisely in this work. The question could be addressed in the future through TA spectroscopy on polymer-SWCNTs in thin-film electrochromic devices.[6]

## CONCLUSION

In summary, we have investigated the charge transfer from photoexcited semiconducting (6,5) SWCNTs to a wide-bandgap wrapping polymer PFO−BPy via femtosecond TA spectroscopy. By spectral and dynamic analysis of the PCT products, we show that the PCT from excited SWCNTs to PFO−BPy can be driven not only by the energetically favorable $E_{33}$ excitation but also by the energetically unfavorable $E_{22}$ excitation under high excitation fluences. The energetically unfavorable PCT originates from the Auger recombination of excitons and charge carriers in the SWCNT, which promotes higher energy excitonic states ($E_{33}$ or higher) and thus makes the charge transfer from the photoexcited narrow-bandgap SWCNT toward the wide-bandgap polymer possible. The spectral dynamics of the SWCNT polaron indicate a time constant of ∼0.9 ps for the interfacial charge separation reaction between the SWCNT and PFO−BPy. The charge recombination may manifest in multiple paths. It includes the direct recombination between the closely associated SWCNT electron polaron and PFO−BPy hole polaron on a time scale of ∼3 ps, while the delocalized SWCNT electron and hole polarons prolong the final charge recombination to the subnanosecond time scale. These findings expand our understanding of the PCT mechanism in Type-I heterojunctions with SWCNTs. When an energy up-conversion process, such as Auger recombination of excitons, takes place in a hybrid system, the energetically unfavorable PCT from a narrow-bandgap nanotube donor to a wide-bandgap polymer acceptor can be driven as well. Such processes might need to be considered for optoelectronic devices that rely on charge separation between nanotubes and semiconducting polymers (e.g., solar cells or photodiodes), in particular in systems where the polymer bandgap is much smaller than that of PFO−BPy.

## ASSOCIATED CONTENT

### Supporting Information

The Supporting Information is available free of charge at https://pubs.acs.org/doi/10.1021/acs.jpcc.0c10171.

Excitonic transitions in the exciton picture, stationary absorption spectra, pump pulse spectra, fluence-dependent NIR transient absorption spectra of SWCNT and SWCNT/PFO−BPy hybrid, transient absorption spectra of PFO−BPy, PFO−BPy polaron in UV−vis transient absorption spectra of the SWCNT/PFO−BPy hybrid and spectroelectrochemistry, elementary excitation analysis of the Auger process, analysis of charge-transfer dynamics, exciton density estimate, and comparison between all-optical doping and chemical doping (PDF)

## AUTHOR INFORMATION

### Corresponding Author

**Tiago Buckup** − *Physikalisch Chemisches Institut and Centre for Advanced Materials, Ruprecht-Karls Universität Heidelberg, Heidelberg 69120, Germany*; orcid.org/0000-0002-1194-0837; Phone: +49 6221 548723; Email: tiago.buckup@pci.uni-heidelberg.de; Fax: +49 6221 548730

### Authors

**Zhuoran Kuang** − *Physikalisch Chemisches Institut and Centre for Advanced Materials, Ruprecht-Karls Universität Heidelberg, Heidelberg 69120, Germany*

**Felix J. Berger** − *Physikalisch Chemisches Institut and Centre for Advanced Materials, Ruprecht-Karls Universität Heidelberg, Heidelberg 69120, Germany*

**Jose Luis Pérez Lustres** − *Physikalisch Chemisches Institut and Centre for Advanced Materials, Ruprecht-Karls Universität Heidelberg, Heidelberg 69120, Germany*

**Nikolaus Wollscheid** − *Physikalisch Chemisches Institut and Centre for Advanced Materials, Ruprecht-Karls Universität Heidelberg, Heidelberg 69120, Germany*

**Han Li** − *Institute of Nanotechnology, Karlsruhe Institute of Technology, Eggenstein-Leopoldshafen 76344, Germany*

**Jan Lüttgens** − *Physikalisch Chemisches Institut and Centre for Advanced Materials, Ruprecht-Karls Universität Heidelberg, Heidelberg 69120, Germany*

**Merve Balcı Leinen** − *Physikalisch Chemisches Institut and Centre for Advanced Materials, Ruprecht-Karls Universität Heidelberg, Heidelberg 69120, Germany*

**Benjamin S. Flavel** − *Institute of Nanotechnology, Karlsruhe Institute of Technology, Eggenstein-Leopoldshafen 76344, Germany;* orcid.org/0000-0002-8213-8673

**Jana Zaumseil** − *Physikalisch Chemisches Institut and Centre for Advanced Materials, Ruprecht-Karls Universität Heidelberg, Heidelberg 69120, Germany;* orcid.org/0000-0002-2048-217X

Complete contact information is available at: https://pubs.acs.org/10.1021/acs.jpcc.0c10171

### Notes

The authors declare no competing financial interest.

## ■ ACKNOWLEDGMENTS

This project received funding from the European Research Council under the European Union's Horizon 2020 research and innovation programme (Grant No. 817494). J.L. acknowledges support by the Volkswagenstiftung (Grant No. 93404). M.B.L. received funding from the Deutsche Forschungsgemeinschaft (DFG, Grant No. ZA 638/7). B.S.F. gratefully acknowledges support from the Deutsche Forschungsgemeinschaft (DFG, Grant No. FL 834/2-2, FL 834/5-1, and FL 834/7-1).

## ■ REFERENCES

(1) Nish, A.; Hwang, J.-Y.; Doig, J.; Nicholas, R. J. Highly selective dispersion of single-walled carbon nanotubes using aromatic polymers. *Nat. Nanotechnol.* **2007**, *2* (10), 640−646.

(2) Graf, A.; Zakharko, Y.; Schießl, S. P.; Backes, C.; Pfohl, M.; Flavel, B. S.; Zaumseil, J. Large scale, selective dispersion of long single-walled carbon nanotubes with high photoluminescence quantum yield by shear force mixing. *Carbon* **2016**, *105*, 593−599.

(3) Samanta, S. K.; Fritsch, M.; Scherf, U.; Gomulya, W.; Bisri, S. Z.; Loi, M. A. Conjugated Polymer-Assisted Dispersion of Single-Wall Carbon Nanotubes: The Power of Polymer Wrapping. *Acc. Chem. Res.* **2014**, *47* (8), 2446−2456.

(4) Rother, M.; Brohmann, M.; Yang, S.; Grimm, S. B.; Schiessl, S. P.; Graf, A.; Zaumseil, J. Aerosol-Jet Printing of Polymer-Sorted (6,5) Carbon Nanotubes for Field-Effect Transistors with High Reproducibility. *Adv. Electron. Mater.* **2017**, *3* (8), 1700080.

(5) Graf, A.; Murawski, C.; Zakharko, Y.; Zaumseil, J.; Gather, M. C. Infrared Organic Light-Emitting Diodes with Carbon Nanotube Emitters. *Adv. Mater.* **2018**, *30* (12), 1706711.

(6) Berger, F. J.; Higgins, T. M.; Rother, M.; Graf, A.; Zakharko, Y.; Allard, S.; Matthiesen, M.; Gotthardt, J. M.; Scherf, U.; Zaumseil, J. From Broadband to Electrochromic Notch Filters with Printed Monochiral Carbon Nanotubes. *ACS Appl. Mater. Interfaces* **2018**, *10* (13), 11135−11142.

(7) Ye, Y.; Bindl, D. J.; Jacobberger, R. M.; Wu, M.-Y.; Roy, S. S.; Arnold, M. S. Semiconducting Carbon Nanotube Aerogel Bulk Heterojunction Solar Cells. *Small* **2014**, *10* (16), 3299−3306.

(8) Pfohl, M.; Glaser, K.; Graf, A.; Mertens, A.; Tune, D. D.; Puerckhauer, T.; Alam, A.; Wei, L.; Chen, Y.; Zaumseil, J.; et al. Probing the Diameter Limit of Single Walled Carbon Nanotubes in SWCNT: Fullerene Solar Cells. *Adv. Energy Mater.* **2016**, *6* (21), 1600890.

(9) Li, G.; Suja, M.; Chen, M.; Bekyarova, E.; Haddon, R. C.; Liu, J.; Itkis, M. E. Visible-Blind UV Photodetector Based on Single-Walled Carbon Nanotube Thin Film/ZnO Vertical Heterostructures. *ACS Appl. Mater. Interfaces* **2017**, *9* (42), 37094−37104.

(10) Schuettfort, T.; Nish, A.; Nicholas, R. J. Observation of a type II heterojunction in a highly ordered polymer-carbon nanotube nanohybrid structure. *Nano Lett.* **2009**, *9* (11), 3871−6.

(11) Eckstein, A.; Karpicz, R.; Augulis, R.; Redeckas, K.; Vengris, M.; Namal, I.; Hertel, T.; Gulbinas, V. Excitation quenching in polyfluorene polymers bound to (6,5) single-wall carbon nanotubes. *Chem. Phys.* **2016**, *467*, 1−5.

(12) Amori, A. R.; Hou, Z.; Krauss, T. D. Excitons in Single-Walled Carbon Nanotubes and Their Dynamics. *Annu. Rev. Phys. Chem.* **2018**, *69*, 81−99.

(13) Bindl, D. J.; Arnold, M. S. Efficient Exciton Relaxation and Charge Generation in Nearly Monochiral (7,5) Carbon Nanotube/C60 Thin-Film Photovoltaics. *J. Phys. Chem. C* **2013**, *117* (5), 2390−2395.

(14) Wang, J.; Peurifoy, S. R.; Bender, M. T.; Ng, F.; Choi, K.-S.; Nuckolls, C.; Arnold, M. S. Non-fullerene Acceptors for Harvesting Excitons from Semiconducting Carbon Nanotubes. *J. Phys. Chem. C* **2019**, *123* (35), 21395−21402.

(15) Mollahosseini, M.; Karunaratne, E.; Gibson, G. N.; Gascon, J. A.; Papadimitrakopoulos, F. Fullerene-Assisted Photoinduced Charge Transfer of Single-Walled Carbon Nanotubes through a Flavin Helix. *J. Am. Chem. Soc.* **2016**, *138* (18), 5904−15.

(16) Balcı Leinen, M.; Berger, F. J.; Klein, P.; Mühlinghaus, M.; Zorn, N. F.; Settele, S.; Allard, S.; Scherf, U.; Zaumseil, J. Doping-Dependent Energy Transfer from Conjugated Polyelectrolytes to (6,5) Single-Walled Carbon Nanotubes. *J. Phys. Chem. C* **2019**, *123* (36), 22680−22689.

(17) Stranks, S. D.; Yong, C.-K.; Alexander-Webber, J. A.; Weisspfennig, C.; Johnston, M. B.; Herz, L. M.; Nicholas, R. J. Nanoengineering Coaxial Carbon Nanotube−Dual-Polymer Heterostructures. *ACS Nano* **2012**, *6* (7), 6058−6066.

(18) Kang, H. S.; Sisto, T. J.; Peurifoy, S.; Arias, D. H.; Zhang, B.; Nuckolls, C.; Blackburn, J. L. Long-Lived Charge Separation at Heterojunctions between Semiconducting Single-Walled Carbon Nanotubes and Perylene Diimide Electron Acceptors. *J. Phys. Chem. C* **2018**, *122* (25), 14150−14161.

(19) Stranks, S. D.; Weisspfennig, C.; Parkinson, P.; Johnston, M. B.; Herz, L. M.; Nicholas, R. J. Ultrafast charge separation at a polymer-single-walled carbon nanotube molecular junction. *Nano Lett.* **2011**, *11* (1), 66−72.

(20) Niklas, J.; Holt, J. M.; Mistry, K.; Rumbles, G.; Blackburn, J. L.; Poluektov, O. G. Charge Separation in P3HT:SWCNT Blends Studied by EPR: Spin Signature of the Photoinduced Charged State in SWCNT. *J. Phys. Chem. Lett.* **2014**, *5* (3), 601−606.

(21) Ferguson, A. J.; Blackburn, J. L.; Holt, J. M.; Kopidakis, N.; Tenent, R. C.; Barnes, T. M.; Heben, M. J.; Rumbles, G. Photoinduced Energy and Charge Transfer in P3HT:SWNT Composites. *J. Phys. Chem. Lett.* **2010**, *1* (15), 2406−2411.

(22) Olivier, J. H.; Park, J.; Deria, P.; Rawson, J.; Bai, Y.; Kumbhar, A. S.; Therien, M. J. Unambiguous Diagnosis of Photoinduced Charge Carrier Signatures in a Stoichiometrically Controlled Semiconducting Polymer-Wrapped Carbon Nanotube Assembly. *Angew. Chem., Int. Ed.* **2015**, *54* (28), 8133−8138.

(23) Kahmann, S.; Salazar Rios, J. M.; Zink, M.; Allard, S.; Scherf, U.; Dos Santos, M. C.; Brabec, C. J.; Loi, M. A. Excited-State Interaction of Semiconducting Single-Walled Carbon Nanotubes with Their Wrapping Polymers. *J. Phys. Chem. Lett.* **2017**, *8* (22), 5666−5672.

(24) Dabera, G. D.; Jayawardena, K. D.; Prabhath, M. R.; Yahya, I.; Tan, Y. Y.; Nismy, N. A.; Shiozawa, H.; Sauer, M.; Ruiz-Soria, G.; Ayala, P.; et al. Hybrid carbon nanotube networks as efficient hole extraction layers for organic photovoltaics. *ACS Nano* **2013**, *7* (1), 556−65.

(25) Deria, P.; Olivier, J. H.; Park, J.; Therien, M. J. Potentiometric, electronic, and transient absorptive spectroscopic properties of oxidized single-walled carbon nanotubes helically wrapped by ionic, semiconducting polymers in aqueous and organic media. *J. Am. Chem. Soc.* **2014**, *136* (40), 14193−9.

(26) Wu, J.; Zhang, S.; Lin, D.; Ma, B.; Yang, L.; Zhang, S.; Kang, L.; Mao, N.; Zhang, N.; Tong, L.; et al. Anisotropic Raman-Enhancement Effect on Single-Walled Carbon Nanotube Arrays. *Adv. Mater. Interfaces* **2018**, *5* (3), 1700941.

(27) Erck, A.; Sapp, W.; Kilina, S.; Kilin, D. Photoinduced Charge Transfer at Interfaces of Carbon Nanotube and Lead Selenide Nanowire. *J. Phys. Chem. C* **2016**, *120* (40), 23197−23206.

(28) Maruyama, S. 1D DOS (van Hove singularity). http://www.photon.t.u-tokyo.ac.jp/~maruyama/kataura/1D_DOS.htmlhttp://www.photon.t.u-tokyo.ac.jp/~maruyama/kataura/1D_DOS.html (accessed March 26, 2020).

(29) Park, K. H.; Lee, S.-H.; Toshimitsu, F.; Lee, J.; Park, S. H.; Tsuyohiko, F.; Jang, J.-W. Gate-enhanced photocurrent of (6,5) single-walled carbon nanotube based field effect transistor. *Carbon* **2018**, *139*, 709−715.

(30) Manzoni, C.; Gambetta, A.; Menna, E.; Meneghetti, M.; Lanzani, G.; Cerullo, G. Intersubband exciton relaxation dynamics in single-walled carbon nanotubes. *Phys. Rev. Lett.* **2005**, *94* (20), 207401.

(31) Crochet, J. J.; Hoseinkhani, S.; Luer, L.; Hertel, T.; Doorn, S. K.; Lanzani, G. Free-carrier generation in aggregates of single-wall carbon nanotubes by photoexcitation in the ultraviolet regime. *Phys. Rev. Lett.* **2011**, *107* (25), 257402.

(32) Park, J.; Reid, O. G.; Blackburn, J. L.; Rumbles, G. Photoinduced spontaneous free-carrier generation in semiconducting single-walled carbon nanotubes. *Nat. Commun.* **2015**, *6*, 8809.

(33) Soavi, G.; Scotognella, F.; Brida, D.; Hefner, T.; Späth, F.; Antognazza, M. R.; Hertel, T.; Lanzani, G.; Cerullo, G. Ultrafast Charge Photogeneration in Semiconducting Carbon Nanotubes. *J. Phys. Chem. C* **2013**, *117* (20), 10849−10855.

(34) Valkunas, L.; Ma, Y.-Z.; Fleming, G. R. Exciton-exciton annihilation in single-walled carbon nanotubes. *Phys. Rev. B: Condens. Matter Mater. Phys.* **2006**, *73* (11), 115432.

(35) Ma, Y. Z.; Valkunas, L.; Dexheimer, S. L.; Bachilo, S. M.; Fleming, G. R. Femtosecond spectroscopy of optical excitations in single-walled carbon nanotubes: evidence for exciton-exciton annihilation. *Phys. Rev. Lett.* **2005**, *94* (15), 157402.

(36) Wang, S.; Khafizov, M.; Tu, X.; Zheng, M.; Krauss, T. D. Multiple exciton generation in single-walled carbon nanotubes. *Nano Lett.* **2010**, *10* (7), 2381−6.

(37) O'Connell, M. J.; Bachilo, S. M.; Huffman, C. B.; Moore, V. C.; Strano, M. S.; Haroz, E. H.; Rialon, K. L.; Boul, P. J.; Noon, W. H.; Kittrell, C.; et al. Band Gap Fluorescence from Individual Single-Walled Carbon Nanotubes. *Science* **2002**, *297* (5581), 593−596.

(38) Bai, Y.; Bullard, G.; Olivier, J. H.; Therien, M. J. Quantitative Evaluation of Optical Free Carrier Generation in Semiconducting Single-Walled Carbon Nanotubes. *J. Am. Chem. Soc.* **2018**, *140* (44), 14619−14626.

(39) Park, J.; Deria, P.; Olivier, J. H.; Therien, M. J. Fluence-dependent singlet exciton dynamics in length-sorted chirality-enriched single-walled carbon nanotubes. *Nano Lett.* **2014**, *14* (2), 504−11.

(40) Huang, L.; Krauss, T. D. Quantized Bimolecular Auger Recombination of Excitons in Single-Walled Carbon Nanotubes. *Phys. Rev. Lett.* **2006**, *96* (5), 057407.

(41) Li, H.; Gordeev, G.; Garrity, O.; Reich, S.; Flavel, B. S. Separation of Small-Diameter Single-Walled Carbon Nanotubes in One to Three Steps with Aqueous Two-Phase Extraction. *ACS Nano* **2019**, *13* (2), 2567−2578.

(42) Pfohl, M.; Tune, D. D.; Graf, A.; Zaumseil, J.; Krupke, R.; Flavel, B. S. Fitting Single-Walled Carbon Nanotube Optical Spectra. *ACS Omega* **2017**, *2* (3), 1163−1171.

(43) Lüer, L.; Hoseinkhani, S.; Polli, D.; Crochet, J.; Hertel, T.; Lanzani, G. Size and mobility of excitons in (6, 5) carbon nanotubes. *Nat. Phys.* **2009**, *5* (1), 54−58.

(44) Styers-Barnett, D. J.; Ellison, S. P.; Mehl, B. P.; Westlake, B. C.; House, R. L.; Park, C.; Wise, K. E.; Papanikolas, J. M. Exciton dynamics and biexciton formation in single-walled carbon nanotubes studied with femtosecond transient absorption spectroscopy. *J. Phys. Chem. C* **2008**, *112* (12), 4507−4516.

(45) Park, J.; Deria, P.; Therien, M. J. Dynamics and transient absorption spectral signatures of the single-wall carbon nanotube electronically excited triplet state. *J. Am. Chem. Soc.* **2011**, *133* (43), 17156−9.

(46) Korovyanko, O. J.; Sheng, C. X.; Vardeny, Z. V.; Dalton, A. B.; Baughman, R. H. Ultrafast spectroscopy of excitons in single-walled carbon nanotubes. *Phys. Rev. Lett.* **2004**, *92* (1), 017403.

(47) Bai, Y.; Olivier, J. H.; Bullard, G.; Liu, C.; Therien, M. J. Dynamics of charged excitons in electronically and morphologically homogeneous single-walled carbon nanotubes. *Proc. Natl. Acad. Sci. U. S. A.* **2018**, *115* (4), 674−679.

(48) Zhu, Z.; Crochet, J.; Arnold, M. S.; Hersam, M. C.; Ulbricht, H.; Resasco, D.; Hertel, T. Pump-Probe Spectroscopy of Exciton Dynamics in (6,5) Carbon Nanotubes. *J. Phys. Chem. C* **2007**, *111* (10), 3831−3835.

(49) Ma, Y.-Z.; Stenger, J.; Zimmermann, J.; Bachilo, S. M.; Smalley, R. E.; Weisman, R. B.; Fleming, G. R. Ultrafast carrier dynamics in single-walled carbon nanotubes probed by femtosecond spectroscopy. *J. Chem. Phys.* **2004**, *120* (7), 3368−3373.

(50) Arias, D. H.; Sulas-Kern, D. B.; Hart, S. M.; Kang, H. S.; Hao, J.; Ihly, R.; Johnson, J. C.; Blackburn, J. L.; Ferguson, A. J. Effect of nanotube coupling on exciton transport in polymer-free monochiral semiconducting carbon nanotube networks. *Nanoscale* **2019**, *11* (44), 21196−21206.

(51) Figueroa Del Valle, D. G.; Moretti, L.; Maqueira-Albo, I.; Aluicio-Sarduy, E.; Kriegel, I.; Lanzani, G.; Scotognella, F. Ultrafast Hole Transfer from (6,5) SWCNT to P3HT:PCBM Blend by Resonant Excitation. *J. Phys. Chem. Lett.* **2016**, *7* (17), 3353−8.

(52) Verissimo-Alves, M.; Capaz, R. B.; Koiller, B.; Artacho, E.; Chacham, H. Polarons in Carbon Nanotubes. *Phys. Rev. Lett.* **2001**, *86* (15), 3372−3375.

(53) Jakubka, F.; Grimm, S. B.; Zakharko, Y.; Gannott, F.; Zaumseil, J. Trion Electroluminescence from Semiconducting Carbon Nanotubes. *ACS Nano* **2014**, *8* (8), 8477−8486.

(54) Matsunaga, R.; Matsuda, K.; Kanemitsu, Y. Observation of charged excitons in hole-doped carbon nanotubes using photoluminescence and absorption spectroscopy. *Phys. Rev. Lett.* **2011**, *106* (3), 037404.

(55) Eckstein, K. H.; Oberndorfer, F.; Achsnich, M. M.; Schöppler, F.; Hertel, T. Quantifying Doping Levels in Carbon Nanotubes by Optical Spectroscopy. *J. Phys. Chem. C* **2019**, *123* (49), 30001−30006.

(56) Heller, I.; Kong, J.; Williams, K. A.; Dekker, C.; Lemay, S. G. Electrochemistry at Single-Walled Carbon Nanotubes: The Role of Band Structure and Quantum Capacitance. *J. Am. Chem. Soc.* **2006**, *128* (22), 7353−7359.

(57) Zheng, M.; Diner, B. A. Solution Redox Chemistry of Carbon Nanotubes. *J. Am. Chem. Soc.* **2004**, *126* (47), 15490−15494.

(58) Eckstein, K. H.; Hartleb, H.; Achsnich, M. M.; Schoppler, F.; Hertel, T. Localized Charges Control Exciton Energetics and Energy Dissipation in Doped Carbon Nanotubes. *ACS Nano* **2017**, *11* (10), 10401−10408.

(59) Hartleb, H.; Späth, F.; Hertel, T. Evidence for Strong Electronic Correlations in the Spectra of Gate-Doped Single-Wall Carbon Nanotubes. *ACS Nano* **2015**, *9* (10), 10461−10470.

(60) Kochi, J. K. Inner-sphere electron transfer in organic chemistry. Relevance to electrophilic aromatic nitration. *Acc. Chem. Res.* **1992**, *25* (1), 39−47.

(61) Connelly, N. G.; Geiger, W. E. Chemical Redox Agents for Organometallic Chemistry. *Chem. Rev.* **1996**, *96* (2), 877−910.

(62) Aguirre, C. M.; Levesque, P. L.; Paillet, M.; Lapointe, F.; St-Antoine, B. C.; Desjardins, P.; Martel, R. The Role of the Oxygen/Water Redox Couple in Suppressing Electron Conduction in Field-Effect Transistors. *Adv. Mater.* **2009**, *21* (30), 3087−3091.

(63) Avery, A. D.; Zhou, B. H.; Lee, J.; Lee, E.-S.; Miller, E. M.; Ihly, R.; Wesenberg, D.; Mistry, K. S.; Guillot, S. L.; Zink, B. L.; et al. Tailored semiconducting carbon nanotube networks with enhanced thermoelectric properties. *Nat. Energy* **2016**, *1* (4), 16033.

(64) Holt, J. M.; Ferguson, A. J.; Kopidakis, N.; Larsen, B. A.; Bult, J.; Rumbles, G.; Blackburn, J. L. Prolonging Charge Separation in P3HT−SWNT Composites Using Highly Enriched Semiconducting Nanotubes. *Nano Lett.* **2010**, *10* (11), 4627−4633.

(65) Sato, K.; Saito, R.; Jiang, J.; Dresselhaus, G.; Dresselhaus, M. S. Discontinuity in the family pattern of single-wall carbon nanotubes. *Phys. Rev. B: Condens. Matter Mater. Phys.* **2007**, *76* (19), 195446.
(66) Chmeliov, J.; Narkeliunas, J.; Graham, M. W.; Fleming, G. R.; Valkunas, L. Exciton−exciton annihilation and relaxation pathways in semiconducting carbon nanotubes. *Nanoscale* **2016**, *8* (3), 1618−1626.
(67) Harrah, D. M.; Schneck, J. R.; Green, A. A.; Hersam, M. C.; Ziegler, L. D.; Swan, A. K. Intensity-Dependent Exciton Dynamics of (6,5) Single-Walled Carbon Nanotubes: Momentum Selection Rules, Diffusion, and Nonlinear Interactions. *ACS Nano* **2011**, *5* (12), 9898−9906.
(68) Perebeinos, V.; Tersoff, J.; Avouris, P. Scaling of Excitons in Carbon Nanotubes. *Phys. Rev. Lett.* **2004**, *92* (25), 257402.
(69) Santos, S. M.; Yuma, B.; Berciaud, S.; Shaver, J.; Gallart, M.; Gilliot, P.; Cognet, L.; Lounis, B. All-Optical Trion Generation in Single-Walled Carbon Nanotubes. *Phys. Rev. Lett.* **2011**, *107* (18), 187401.
(70) Soavi, G.; Scotognella, F.; Viola, D.; Hefner, T.; Hertel, T.; Cerullo, G.; Lanzani, G. High energetic excitons in carbon nanotubes directly probe charge-carriers. *Sci. Rep.* **2015**, *5*, 9681.
(71) Shi, J.; Chu, H.; Li, Y.; Zhang, X.; Pan, H.; Li, D. Synthesis and nonlinear optical properties of semiconducting single-walled carbon nanotubes at 1 μm. *Nanoscale* **2019**, *11* (15), 7287−7292.
(72) Brady, G. J.; Joo, Y.; Singha Roy, S.; Gopalan, P.; Arnold, M. S. High performance transistors via aligned polyfluorene-sorted carbon nanotubes. *Appl. Phys. Lett.* **2014**, *104* (8), 083107.
(73) Ehli, C.; Oelsner, C.; Guldi, D. M.; Mateo-Alonso, A.; Prato, M.; Schmidt, C.; Backes, C.; Hauke, F.; Hirsch, A. Manipulating single-wall carbon nanotubes by chemical doping and charge transfer with perylene dyes. *Nat. Chem.* **2009**, *1* (3), 243−9.
(74) Dowgiallo, A.-M.; Mistry, K. S.; Johnson, J. C.; Blackburn, J. L. Ultrafast Spectroscopic Signature of Charge Transfer between Single-Walled Carbon Nanotubes and $C_{60}$. *ACS Nano* **2014**, *8* (8), 8573−8581.
(75) Howard, I. A.; Mauer, R.; Meister, M.; Laquai, F. Effect of Morphology on Ultrafast Free Carrier Generation in Polythiophene:-Fullerene Organic Solar Cells. *J. Am. Chem. Soc.* **2010**, *132* (42), 14866−14876.

# Supporting Information

# Charge Transfer from Photo-Excited Semiconducting Single-Walled Carbon Nanotubes to Wide-Bandgap Wrapping Polymer

Zhuoran Kuang,[1,2] Felix J. Berger,[1,2] Jose Luis Pérez Lustres,[1,2,□] Nikolaus Wollscheid,[1,2] Han Li,[3] Jan Lüttgens,[1,2] Merve Balcı Leinen,[1,2] Benjamin S. Flavel,[3] Jana Zaumseil,[1,2] Tiago Buckup[1,2,*]

[1] Physikalisch Chemisches Institut, Ruprecht-Karls Universität Heidelberg, Im Neuenheimer Feld 229/253, Heidelberg, 69120, Germany.

[2] Centre for Advanced Materials, Ruprecht-Karls Universität Heidelberg, Im Neuenheimer Feld 225, Heidelberg, 69120, Germany.

[3] Institute of Nanotechnology, Karlsruhe Institute of Technology, Eggenstein-Leopoldshafen, 76344, Germany.

□ Current Address: Institut für Experimentelle Physik, Freie Universität Berlin, Arnimallee 14, Berlin, 14195, Germany

*<u>Corresponding Author</u>:

Dr. Tiago Buckup, tiago.buckup@pci.uni-heidelberg.de, Phone: +49 6221 548723, Fax: +49 6221 548730

# Contents

## A. Excitonic Transitions in Exciton Picture

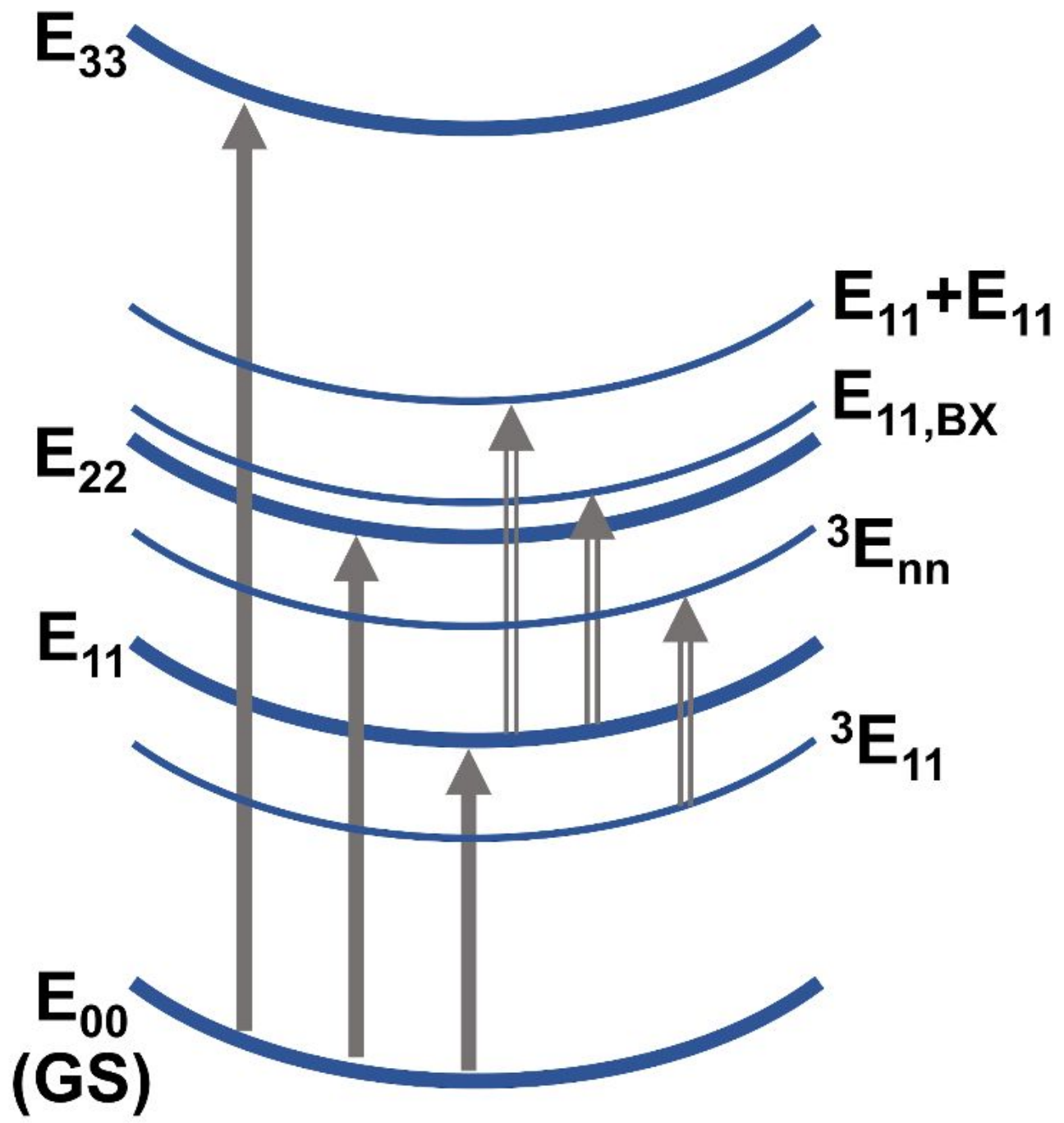

**Figure S1.** Schematic representation of various excitonic transitions in SWCNTs shown in exciton picture. The $E_{00}$ denotes the ground excitonic state (GS) in exciton picture. The solid arrows represent the ground-state absorption, and the hollow arrows represent the excited-state absorption.

## B. Stationary Absorption Spectra, Pump Pulse Spectra, Sample Purities

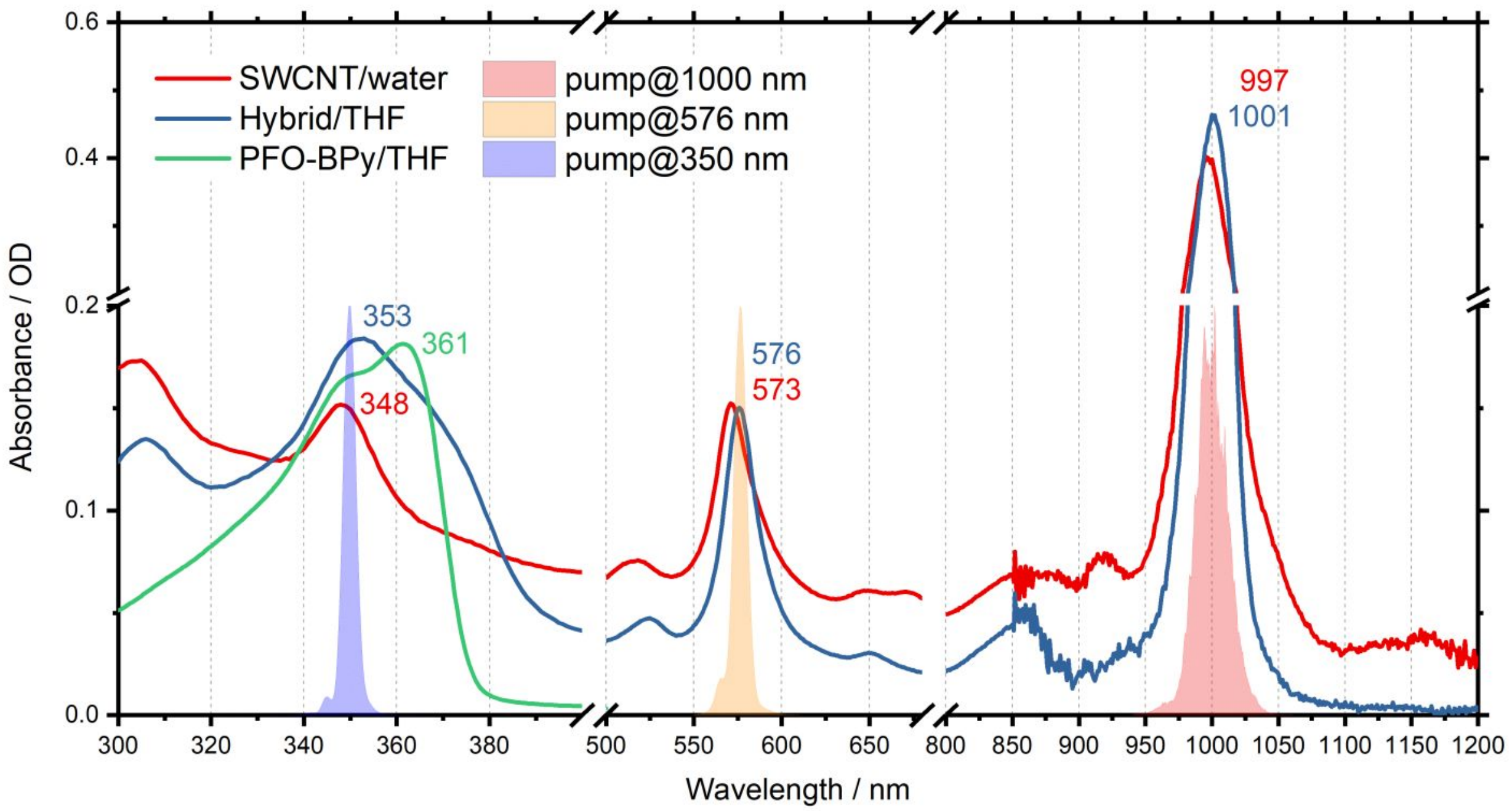

**Figure S2.** Stationary absorption spectra of surfactant-dispersed (6,5) **SWCNT** in water, PFO-BPy-wrapped (6,5) SWCNT **Hybrid** in THF, and PFO-BPy in THF. The positions of major absorption peaks are marked with corresponding colors. Spectra of pump pulse centered at 1000 nm, 576 nm, and 350 nm are plotted in shaded areas.

Even SWCNT dispersions that are highly enriched in a single chirality still contain minority chiralities with an abundance of a few percent. In order to identify these species in our samples and quantify their abundances, we fitted the steady-state absorption spectra displayed in Figure 2 of the main text using the code reported by Pfohl et al.[1] All chiralities with an abundance ≥1 % that were identified in the **SWCNT** or **Hybrid** sample are listed in the Tables S1 and S2.

**Table S1.** Chirality distribution of the sample "SWCNT", *i.e*. ATPE-sorted (6,5) SWCNTs in water.

| Chirality | (6,5) | (7,5) | (8,3) | (9,1) | (9,2) | (7,6) | (8,7) | (9,4) | (8,4) |
|---|---|---|---|---|---|---|---|---|---|
| $\lambda_{11}$ / nm | 997 | 1023 | 972 | 920 | 1163 | 1133 | 1287 | 1106 | 1203 |
| Abundance / % | 62.8 | 13.2 | 6.7 | 4.6 | 4.3 | 2.0 | 1.4 | 1.3 | 1.2 |

**Table S2.** Chirality distribution of the sample "Hybrid", *i.e*. PFO-BPy-wrapped (6,5) SWCNTs in THF.

| Chirality | (6,5) | (8,3) | (9,1) | (5,4) | (7,5) |
|---|---|---|---|---|---|
| $\lambda_{11}$ / nm | 1000 | 972 | 936 | 856 | 1023 |
| Abundance / % | 85.0 | 7.7 | 3.0 | 2.0 | 2.0 |

In agreement with other reports[2], we observe that the PFO-BPy-sorted (6,5) SWCNTs (**Hybrid**) feature a higher purity than a typical ATPE-sorted (**SWCNT**) sample, as evident from the smaller number of chiralities with an abundance ≥1 %. (7,5) SWCNTs are particularly relevant as their $E_{11}$ absorption partially overlaps with the suspected (6,5) SWCNT polaron signal and their concentration in the **SWCNT** sample is significantly higher than in the **Hybrid** sample, because of the different sorting methods employed. However, given the narrow pump spectra (Figure S2) that resonantly excite the (6,5) SWCNTs, the contribution from (7,5) SWCNTs should be negligible.

## C. Fluence-Dependent NIR Transient Absorption Spectra of SWCNT and SWCNT/PFO-BPy Hybrid

Excitation-fluence-dependent transient absorption (TA) measurements enable us to clearly examine the charge-transfer transient products in the SWCNT/PFO-BPy **Hybrid** system. In addition to the results of **Hybrid** presented in the main text, here we provide the fluence-dependent TA spectra for **SWCNT** in water as a reference. The selected TA spectra of **SWCNT** upon $E_{11}$, $E_{22}$, and $E_{33}$ excitation are shown in Figure S3, S4, and S5. In stark contrast with the TA spectra of **Hybrid** (Figure 7d,e,f in the main text), the normalized TA spectra of **SWCNT** at the time delay of 3 ps (Figure S6) do not manifest obvious spectroscopic feature of (6,5) SWCNT polaron. The $E_{00} \rightarrow E_{11}$ transition bleach in spectra does not shift (also shown in Figure 5 in the main text), and polaron absorption around 1050 nm does not exist.

The supplementary fluence-dependent TA spectra of **Hybrid** upon the $E_{11}$, $E_{22}$, and $E_{33}$ excitation are shown in Figure S7, S8, and S9, respectively.

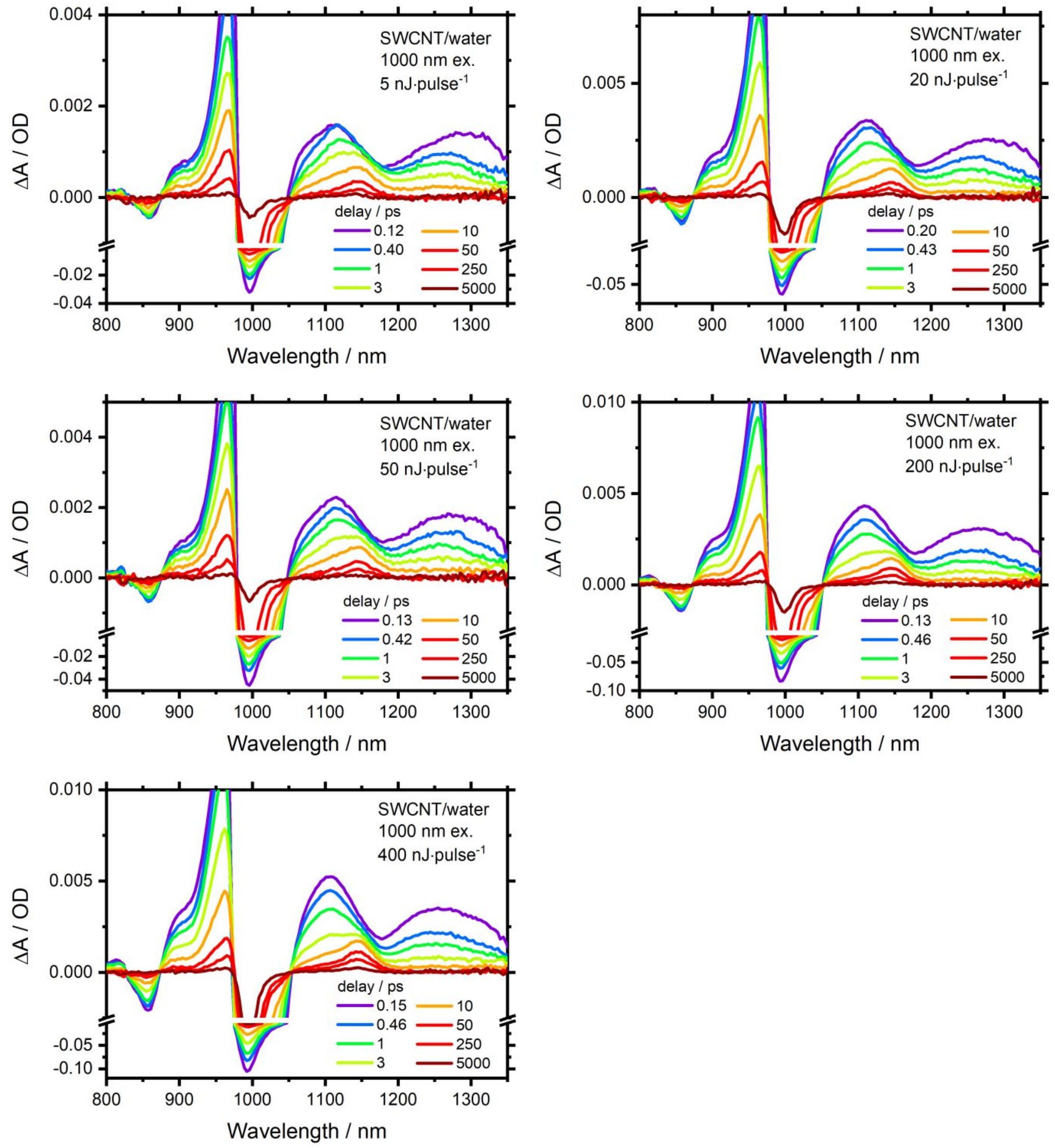

**Figure S3.** Fluence-dependent TA spectra for **SWCNT** in water upon the $E_{11}$ excitation. Excitation wavelength: 1000 nm. Pump-energy-per-pulse is indicated in the legend.

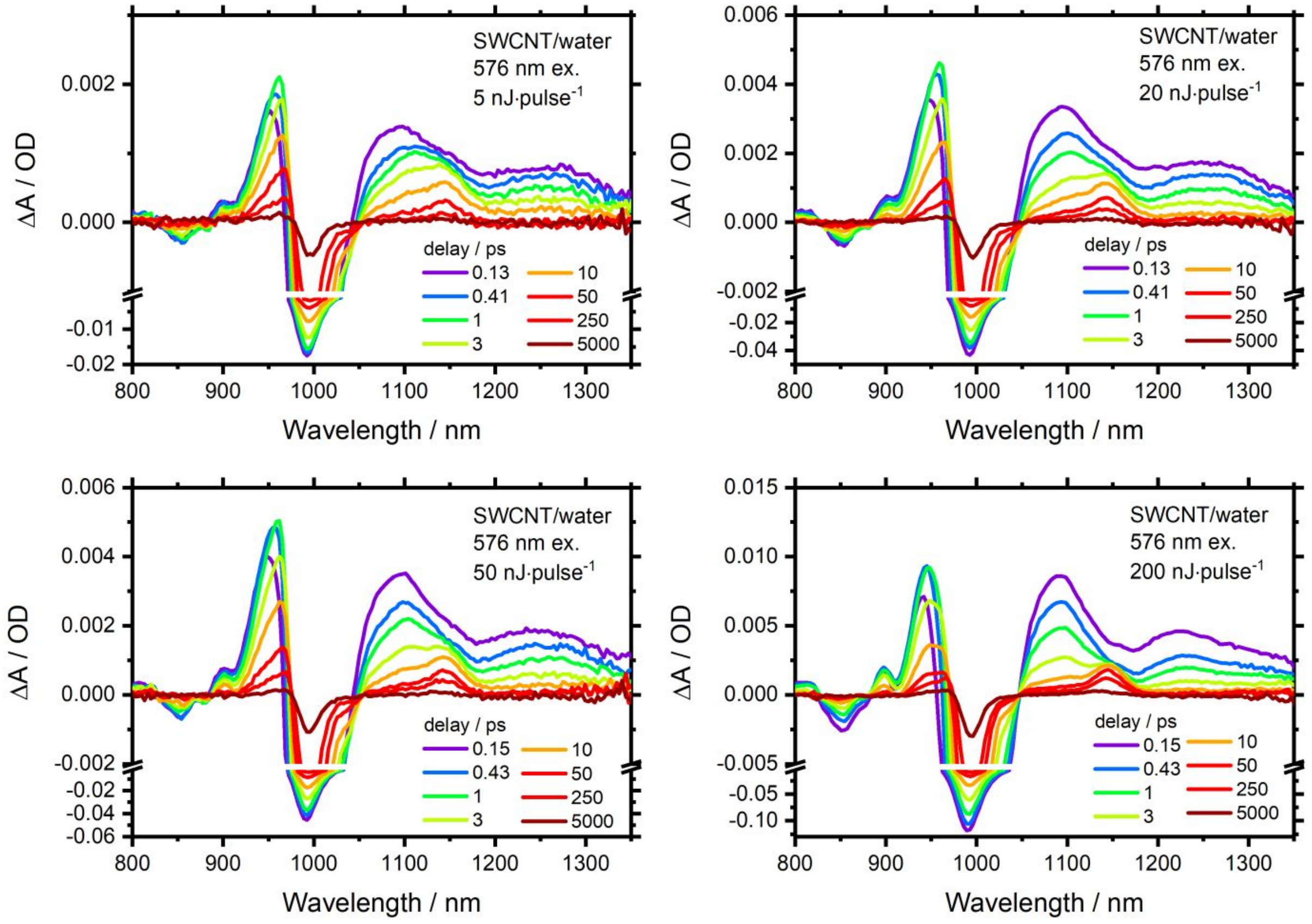

**Figure S4.** Fluence-dependent TA spectra for **SWCNT** in water upon the $E_{22}$ excitation. Excitation wavelength: 576 nm. Pump-energy-per-pulse is indicated in the legend.

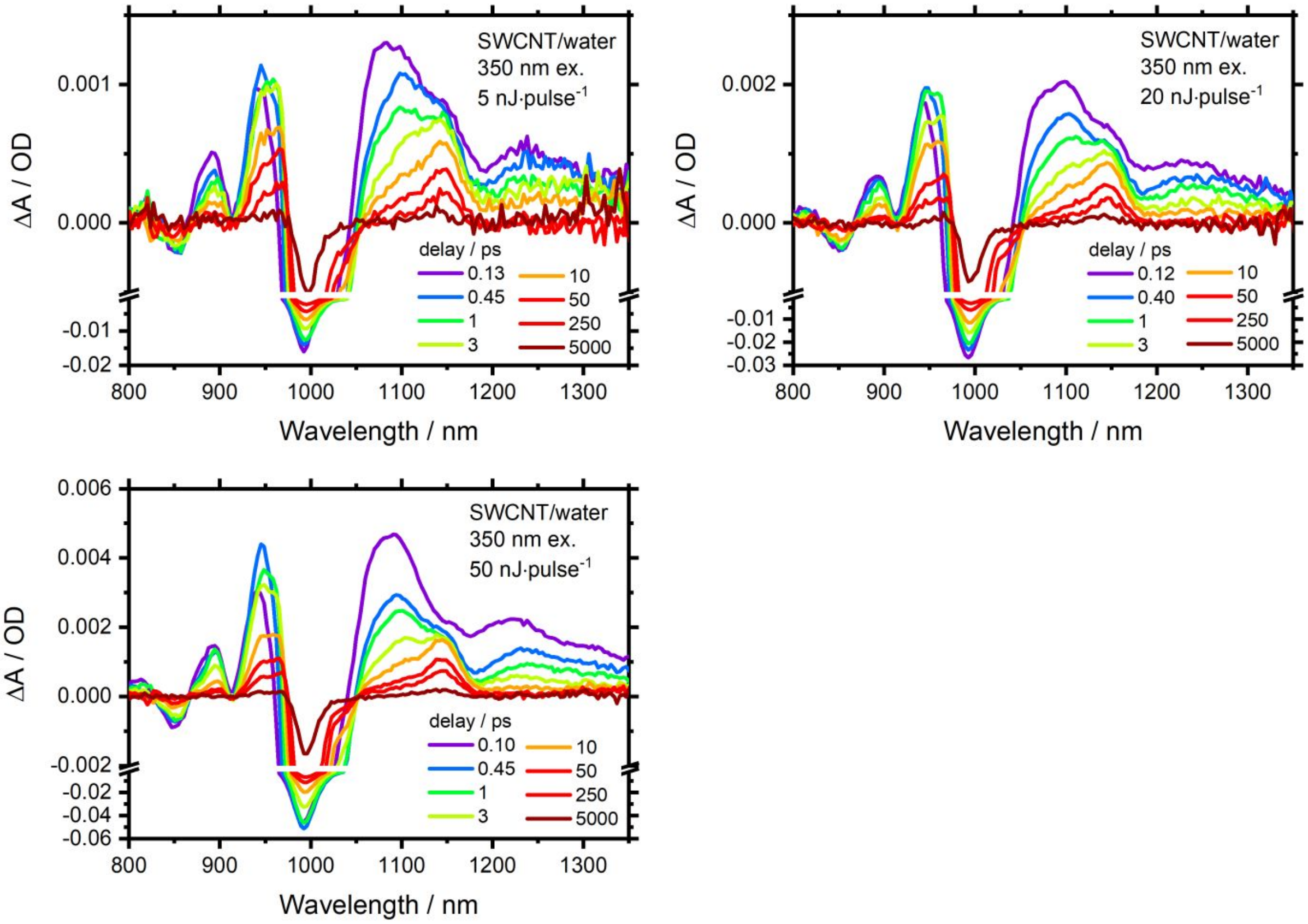

**Figure S5.** Fluence-dependent TA spectra for **SWCNT** in water upon the $E_{33}$ excitation. Excitation wavelength: 350 nm. Pump-energy-per-pulse is indicated in the legend.

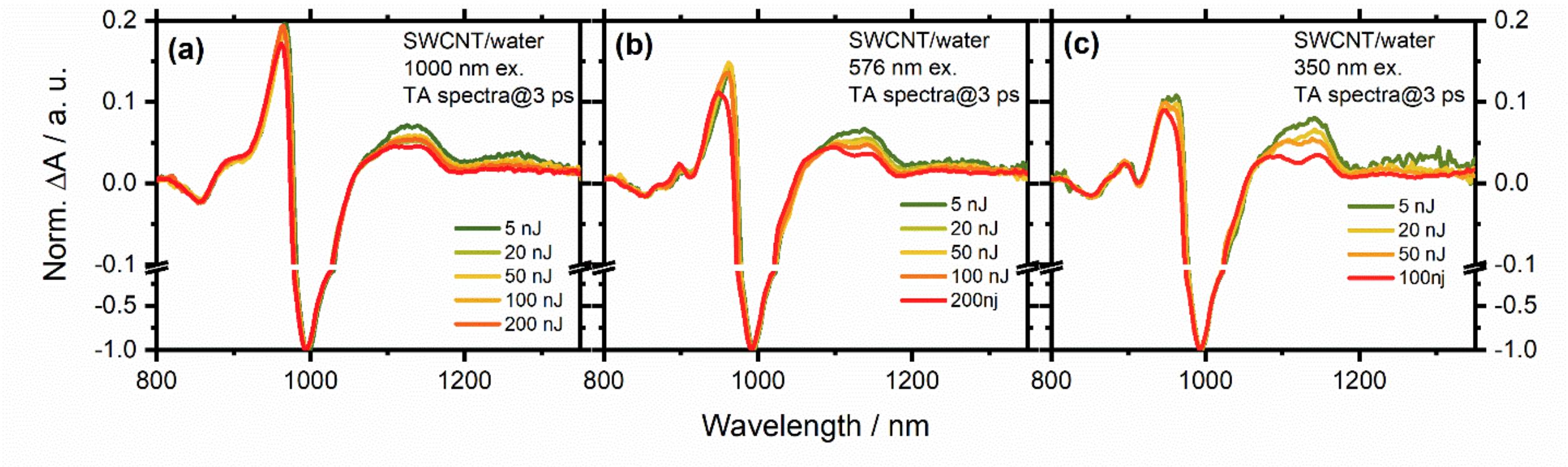

**Figure S6.** Normalized pump-energy-dependent TA spectra for **SWCNT** in water at a time delay of ~3 ps upon (a) $E_{11}$, (b) $E_{22}$, and (c) $E_{33}$ excitation. Note that spectra were normalized at the $E_{11}$ bleaching maximum. Sample information, excitation wavelength, and curves corresponding pump-energy-per-pulse are depicted in legends.

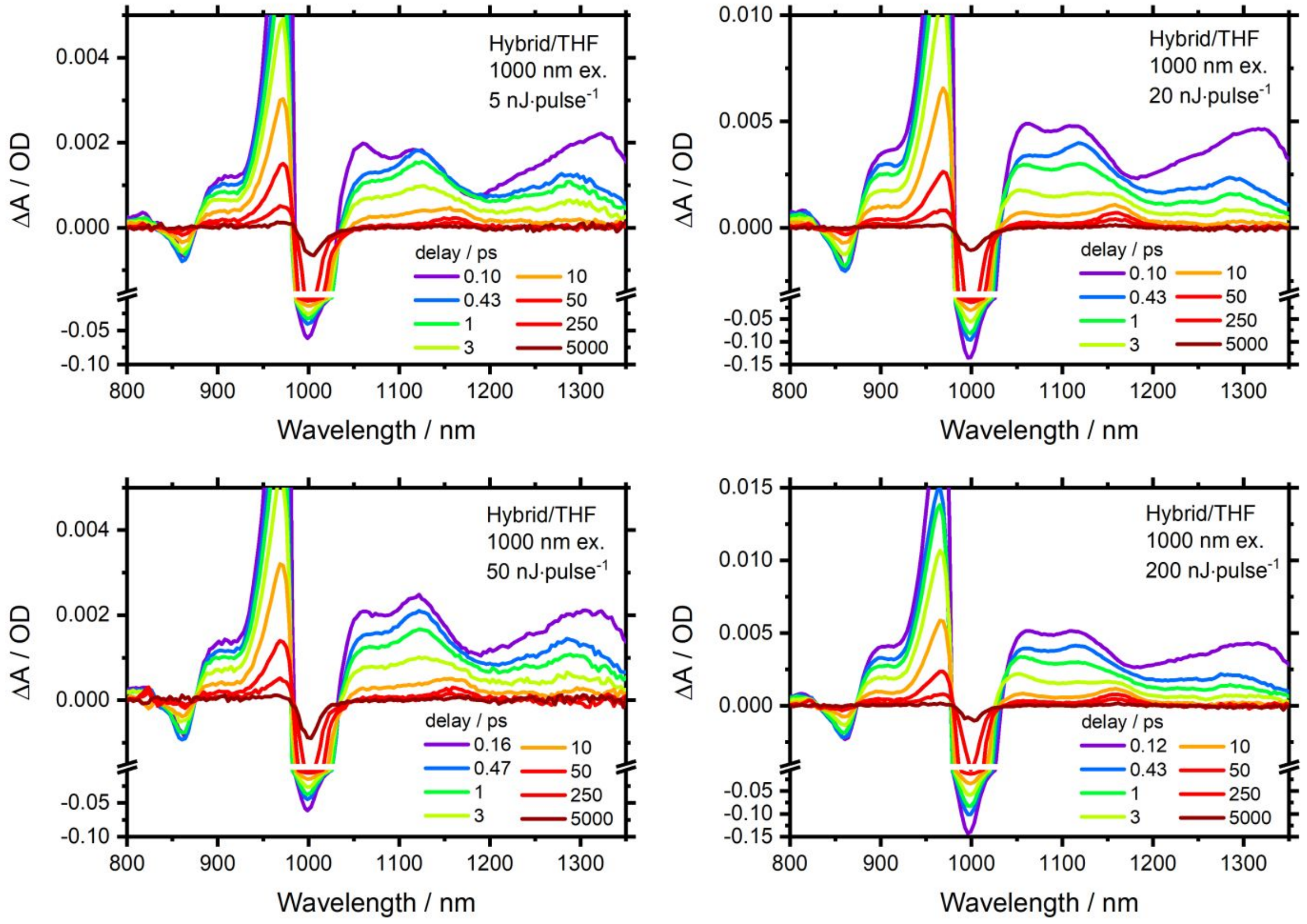

**Figure S7.** Fluence-dependent TA spectra for **Hybrid** in THF upon the $E_{11}$ excitation. Excitation wavelength: 1000 nm. Pump-energy-per-pulse is indicated in the legend.

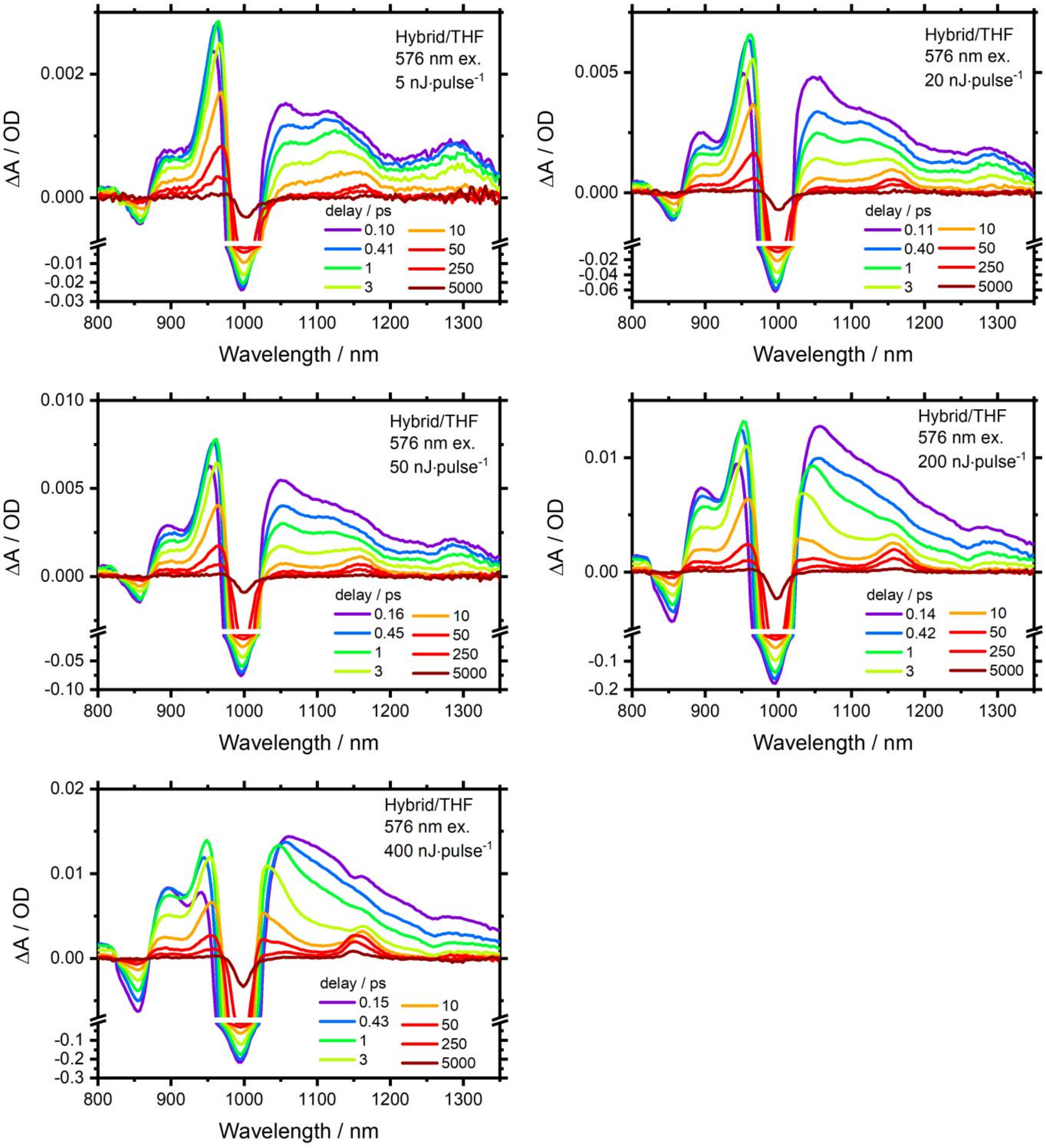

**Figure S8.** Fluence-dependent TA spectra for **Hybrid** in THF upon the $E_{22}$ excitation. Excitation wavelength: 576 nm. Pump-energy-per-pulse is indicated in the legend.

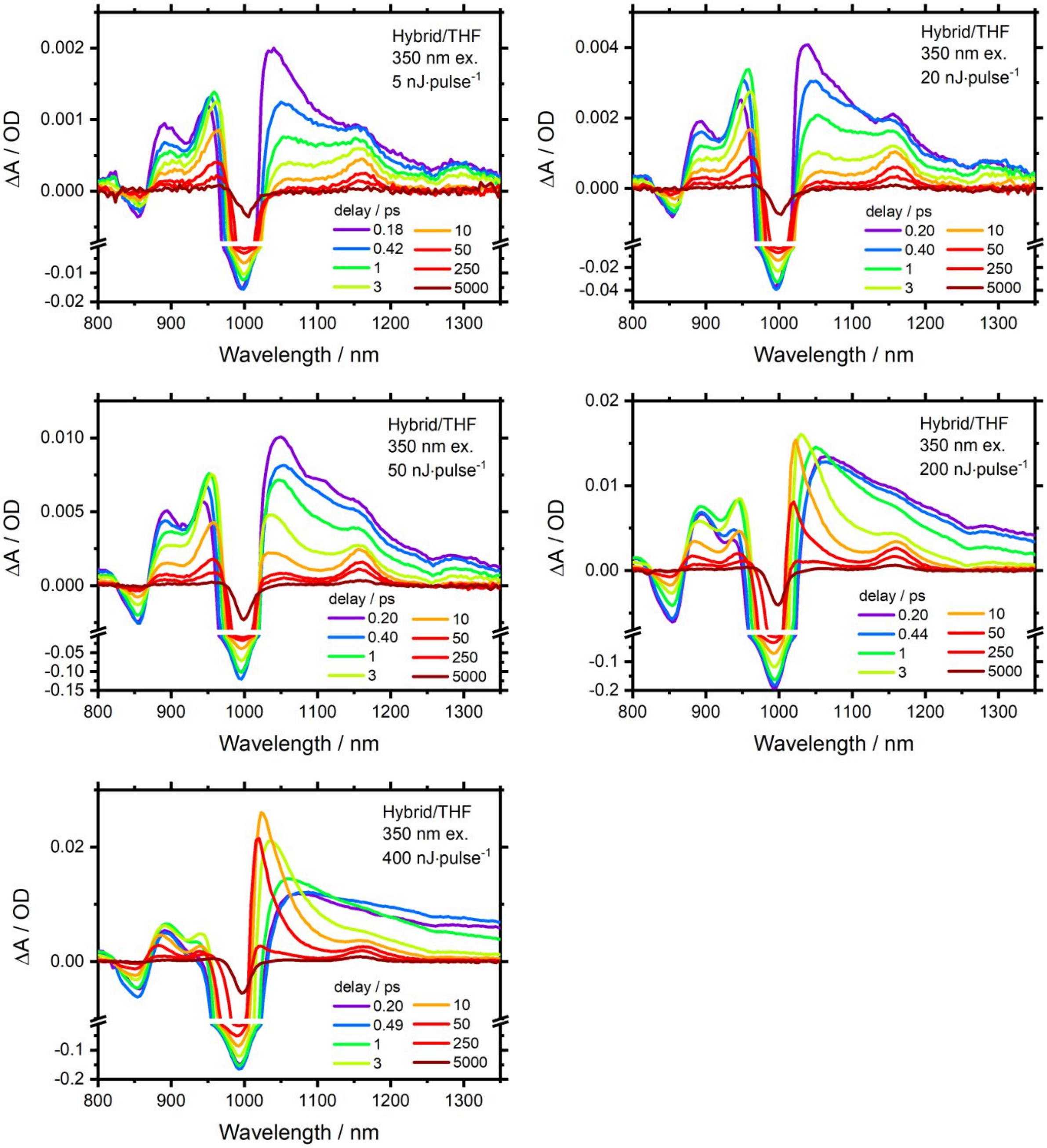

**Figure S9.** Fluence-dependent TA spectra for **Hybrid** in THF upon the $E_{33}$ excitation. Excitation wavelength: 350 nm. Pump-energy-per-pulse is indicated in the legend.

## D. Transient Absorption Spectra of PFO-BPy

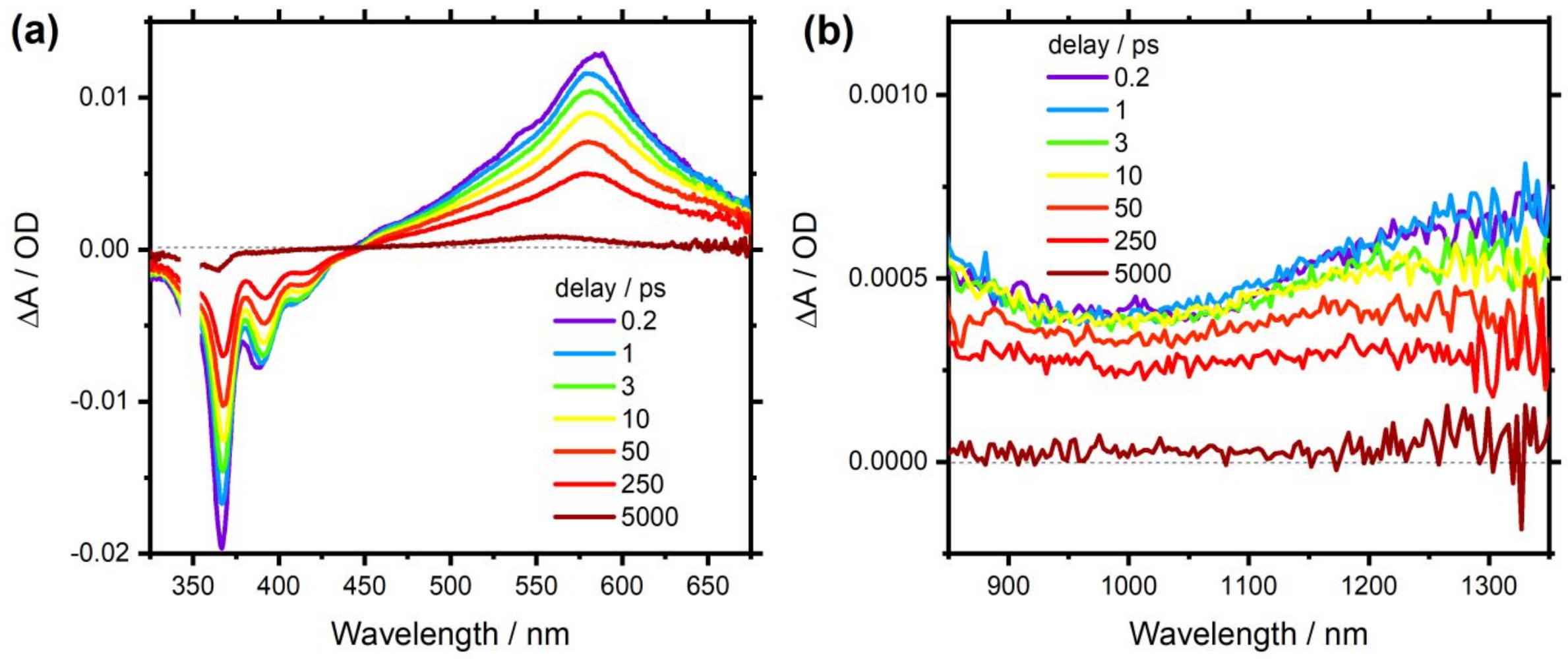

**Figure S10.** Selected TA spectra for PFO-BPy in THF in the (a) UV-Vis and (b) NIR region. Excitation wavelength: 350 nm. Pump energy: 100 nJ·pulse$^{-1}$. Note that the UV-Vis and NIR TA spectra were acquired from independent measurements.

TA spectra of PFO-BPy in THF were measured from UV to NIR region upon 350 nm resonant excitation (Figure S10). In the UV region, the dominated negative absorptive band around 365 nm, which overlaps with its stationary optical absorption, originates from the photo-induced bleach of ground-state absorption. The negative-going absorptive signal between 380 – 450 matches well with the reported stationary fluorescence spectrum of PFO-BPy,[3] thus it should be attributed to the stimulated emission of the lowest excitonic state. The photo-induced excitonic absorption of PFO-BPy, of which the lineshape is broad and structureless, peaks around 580 nm and extends to the NIR region with a very low amplitude.[3]

To exclude the spectral contribution of PFO-BPy when analyzing the SWCNT polaron kinetics of **Hybrid** in NIR region upon the $E_{33}$ excitation, we examine the kinetic traces of PFO-BPy in the spectral region that overlaps with the suspected (6,5) SWCNT polaron between 1000 to 1150 nm. The excitonic absorption band decays evenly in this region which can be adequately fitted with a mono-exponential lifetime of ~660 ps. The fitting results are shown in Figure S11.

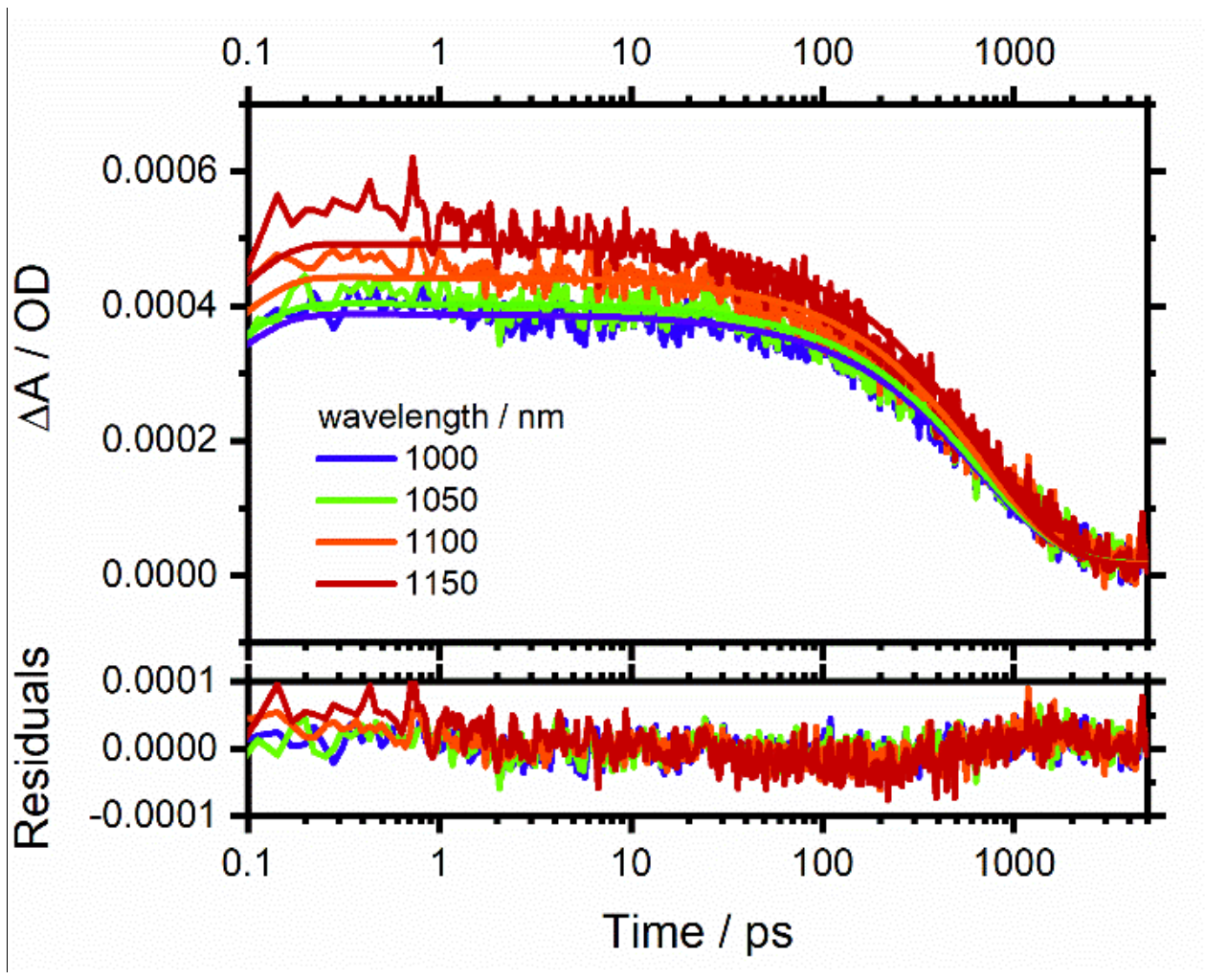

**Figure S11.** Selected transient absorption kinetics trace and fitting results for PFO-BPy in THF at the wavelength of 1000, 1050, 1100, and 1150 nm. The upper panel shows the raw data points and the fitting curves; the lower panel shows the fitting residuals. Excitation wavelength: 350 nm. Pump energy: 100 nJ·pulse$^{-1}$.

## E. PFO-BPy Polaron in UV-Vis Transient Absorption Spectra of SWCNT/PFO-BPy Hybrid

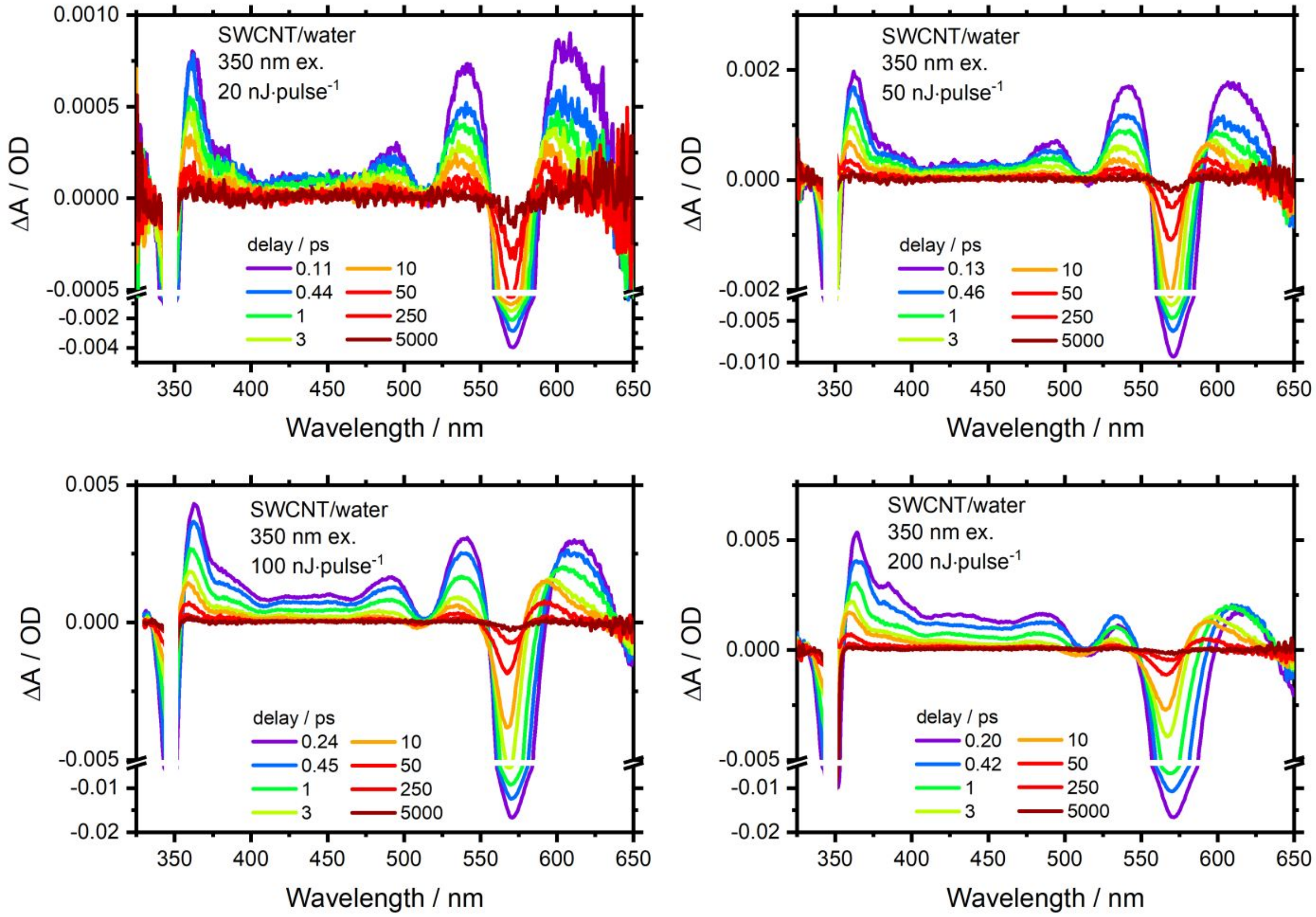

**Figure S12.** Fluence-dependent UV-Vis TA spectra for **SWCNT** in water upon the $E_{33}$ excitation. Excitation wavelength: 350 nm. Pump-energy-per-pulse is indicated in the legend.

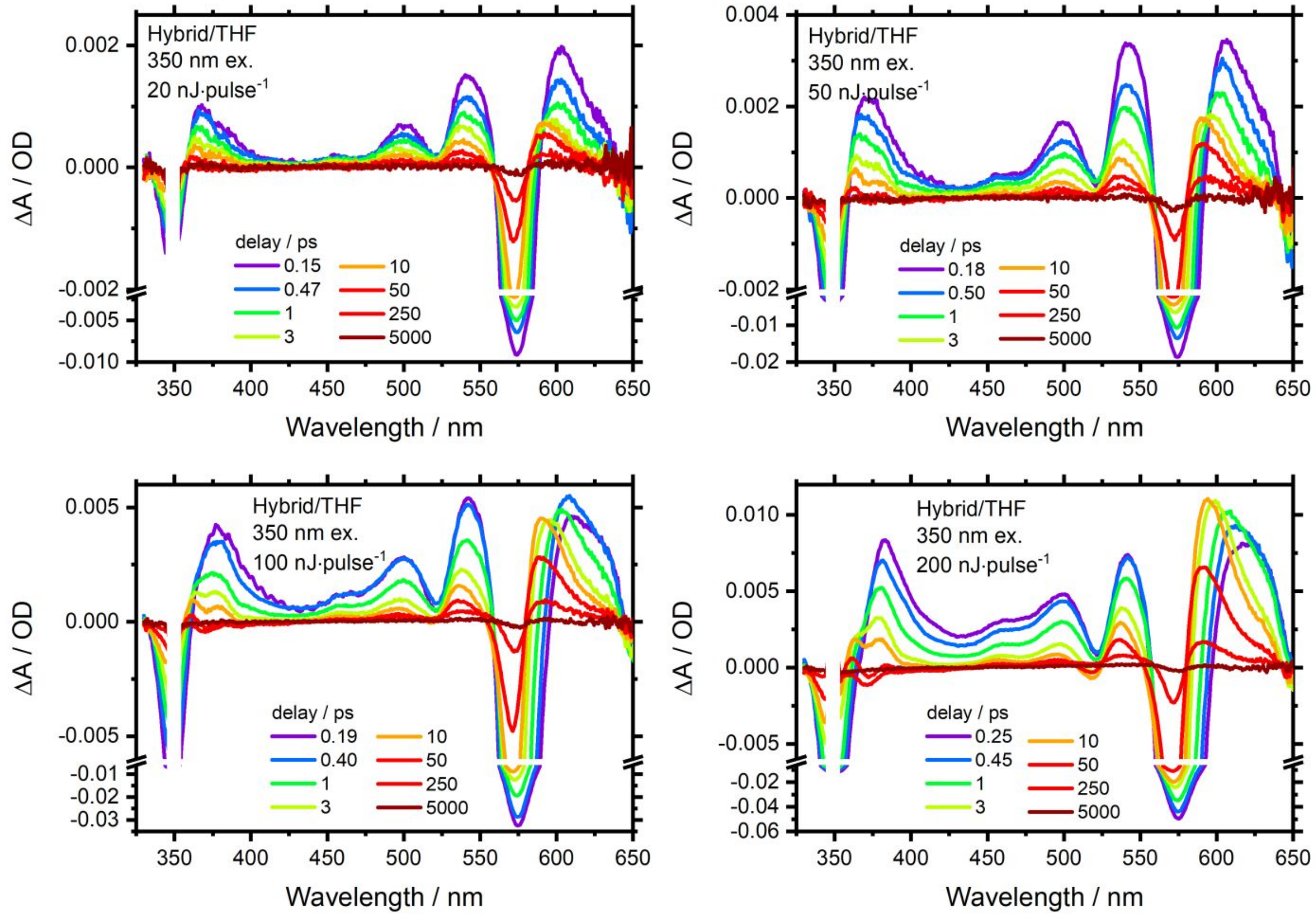

**Figure S13.** Fluence-dependent UV-Vis TA spectra for Hybrid in THF upon the $E_{33}$ excitation. Excitation wavelength: 350 nm. Pump-energy-per-pulse is indicated in the legend.

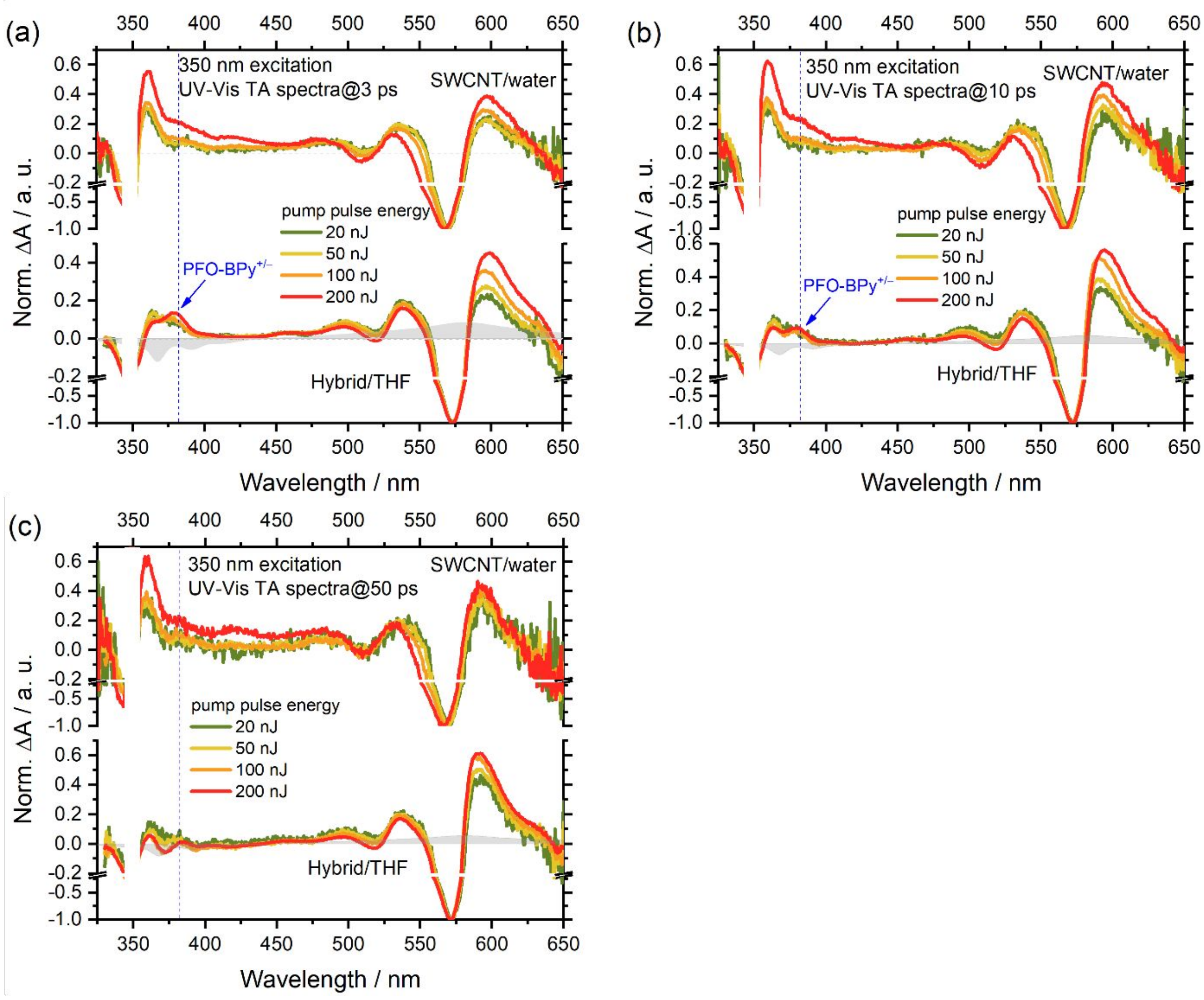

**Figure S14.** Normalized fluence-dependent UV-Vis TA spectra for **SWCNT** (upper panel) and **Hybrid** (lower panel) at the time delay of (a) 3 ps, (b) 10 ps, and (c) 50 ps upon the $E_{33}$ excitation. Note that spectra were normalized at the $E_{22}$ bleaching maximum. Excitation wavelength: 350 nm. Curves corresponding pump-energy-per-pulse were depicted in legends. The gray shaded area in the lower panel represents the TA spectrum of PFO-BPy in THF at the time delay of (a) 3 ps, (b) 10 ps, and (c) 50 ps upon 350 nm excitation. The amplitude of the spectrum is adjusted accordingly for a better comparison with the normalized spectra of **Hybrid**.

For the photo-induced charge transfer (PCT) in SWCNT/PFO-BPy **Hybrid** system, if the formation of SWCNT polaron occurred upon photo-excitation, the transient absorptive signature of transient PFO-BPy polaron should emerge concomitantly. The fluence-dependent UV-Vis TA spectra for **SWCNT** and **Hybrid** upon the $E_{33}$ excitation (350 nm, 3.54 eV) were measured and shown in Figure S12 and S13. For the case of 350 nm excitation, the (6,5) SWCNT and PFO-BPy in **Hybrid** are both resonantly excited. When excited-state interaction between (6,5) SWCNT and PFO-BPy is not considered, the obtained TA spectra of **Hybrid** should be approximately regarded as a linear superposition of TA spectra (6,5) of SWCNT and PFO-BPy. Because of the complex excitonic transition of (6,5) SWCNT (as shown in Figure S2), the TA spectra of **SWCNT** and **Hybrid** become more structured in the UV-Vis region. However, the TA spectra of **Hybrid** manifest an additional absorptive signature around 380 nm at time delays around 1 – 10 ps in comparison with the spectra of **SWCNT**.

Considering that the absorption amplitude of (6,5) SWCNT polaron reaches its maximum at ~3 ps, the fluence-dependent TA spectra of **SWCNT** and **Hybrid**, as well as the spectrum of pure PFO-BPy, are carefully compared in Figure S14a. The TA spectra of **Hybrid** in the spectral region of 450 – 650 nm manifest a simple superposition of SWCNT spectra and positive exciton absorption of PFO-BPy, while in the region of 330 – 450 nm, a positive-going absorptive signal around 380 nm emerges in the spectra of **Hybrid** along with the increased pump fluence. The TA spectrum of PFO-BPy at ~3 ps, or through the entire measurement time window (as shown in Figure S9), does not show any positive contribution in ΔA in the region of 330 – 450 nm. Therefore, the new absorption band around 380 nm should be attributed to the PFO-BPy polaron (PFO-$BPy^{+/-}$). With the same method, the fluence-dependent TA spectra of **SWCNT** and **Hybrid** at 10 ps and 50 ps were compared as well in Figure S14b. The absorptive feature of the PFO-BPy polaron becomes very weak at 10 ps, and hardly observable at 50 ps, even at the highest excitation energy. These results indicate that the timescale of PFO-BPy polaron decay in photo-excited **Hybrid** system is around a few picoseconds, which is obviously shorter than that of the SWCNT polaron. Dynamics of SWCNT polaron refers to the main text, Section *Charge Separation and Recombination Dynamics*, and Supporting Information, Section H.

## F. PFO-BPy Polaron in Spectroelectrochemistry

As discussed in the main text, we deduced that the photo-induced electron-transfer from SWCNT to PFO-BPy is more likely to take place. The photo-induced charge-separated products in the charge-neutral Hybrid system should be the SWCNT hole-polaron and PFO-BPy electron-polaron. Due to the high LUMO energy of the PFO-BPy, it is difficult to prepare its electron-polaron with common reducing agents in a redox titration. In order to support the hypothesis of the PFO-BPy electron-polaron, we carry out spectroelectrochemical measurements for PFO-BPy in THF. After a reduction potential of $-2$V relative to the reference electrode Ag/AgCl was applied on the solution, a new absorption band above 370 nm formed within a few minutes (Figure S15a), while the main absorption band of PFO-BPy around 360 nm dropped simultaneously. After around 20 minutes, the spectral shape stabilized. Taking the spectrum at 24 min as the reduced product spectrum, we obtain the PFO-BPy electron-polaron differential absorption spectrum. As shown in Figure S15b, the differential spectrum of PFO-BPy electron-polaron has the absorption peak around 385 nm. This result matches well with the PCT induced PFO-BPy electron polaron observed in the UV-visible TA spectra of PFO-BPy-wrapped (6,5) SWCNTs.

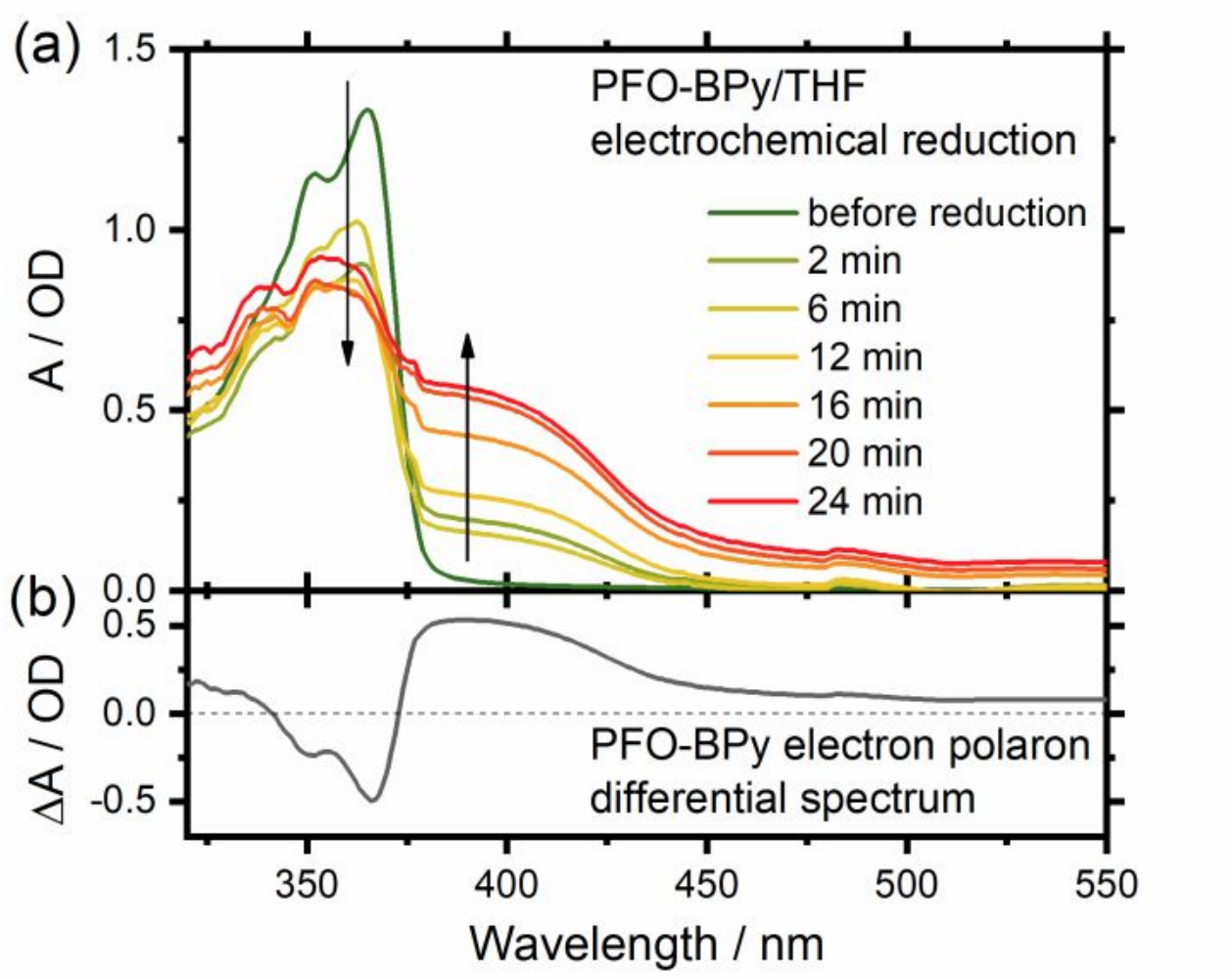

**Figure S15.** (a) UV-Vis absorption spectra in electrochemical reduction measurement of PFO-BPy in THF at an applied potential of -2V. The reaction time is shown in the legends. The reference electrode is Ag/AgCl. (b) Differential spectrum of the PFO-Bpy electron polaron.

## G. Elementary Excitation Analysis of Auger Process in SWCNT/PFO-BPy Hybrid

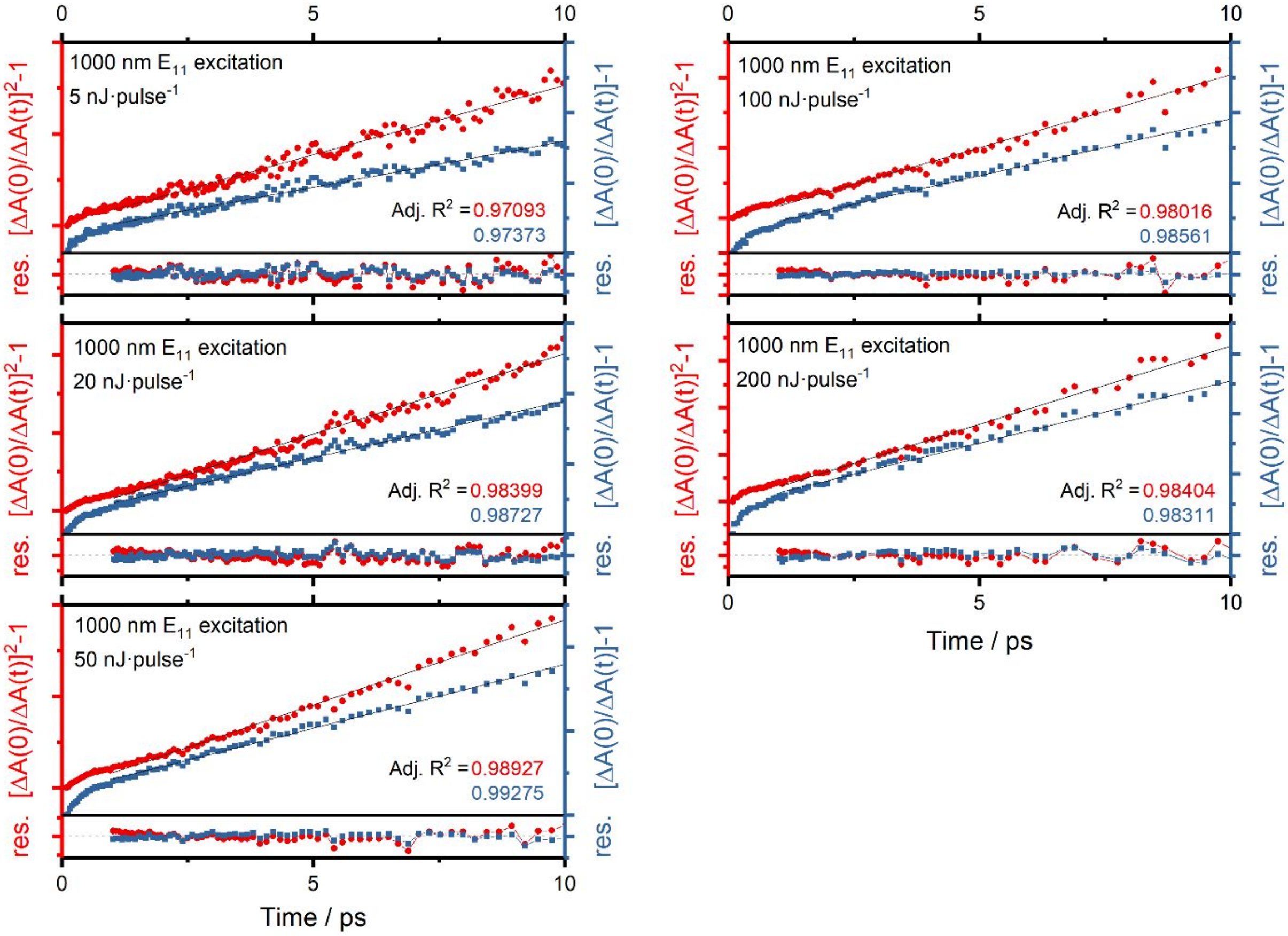

**Figure S16.** Kinetics of the integral $E_{00} \rightarrow E_{11}$ bleaching in the TA spectra of **Hybrid** in THF upon the $E_{11}$ excitations, plotted as {$[\Delta A(0)/\Delta A(t)]^2-1$} (red dots, left axis) and {$[\Delta A(0)/\Delta A(t)]-1$} (blue squares, right axis). Traces are shifted by different offset on the vertical axis for a better comparison. Solid black lines represent the results of the linear fitting. Adjusted R-squared and fitting residuals are shown with corresponding colors. Pump energy: 5, 20, 50, 100, and 200 nJ·pulse$^{-1}$, as indicated in the legends.

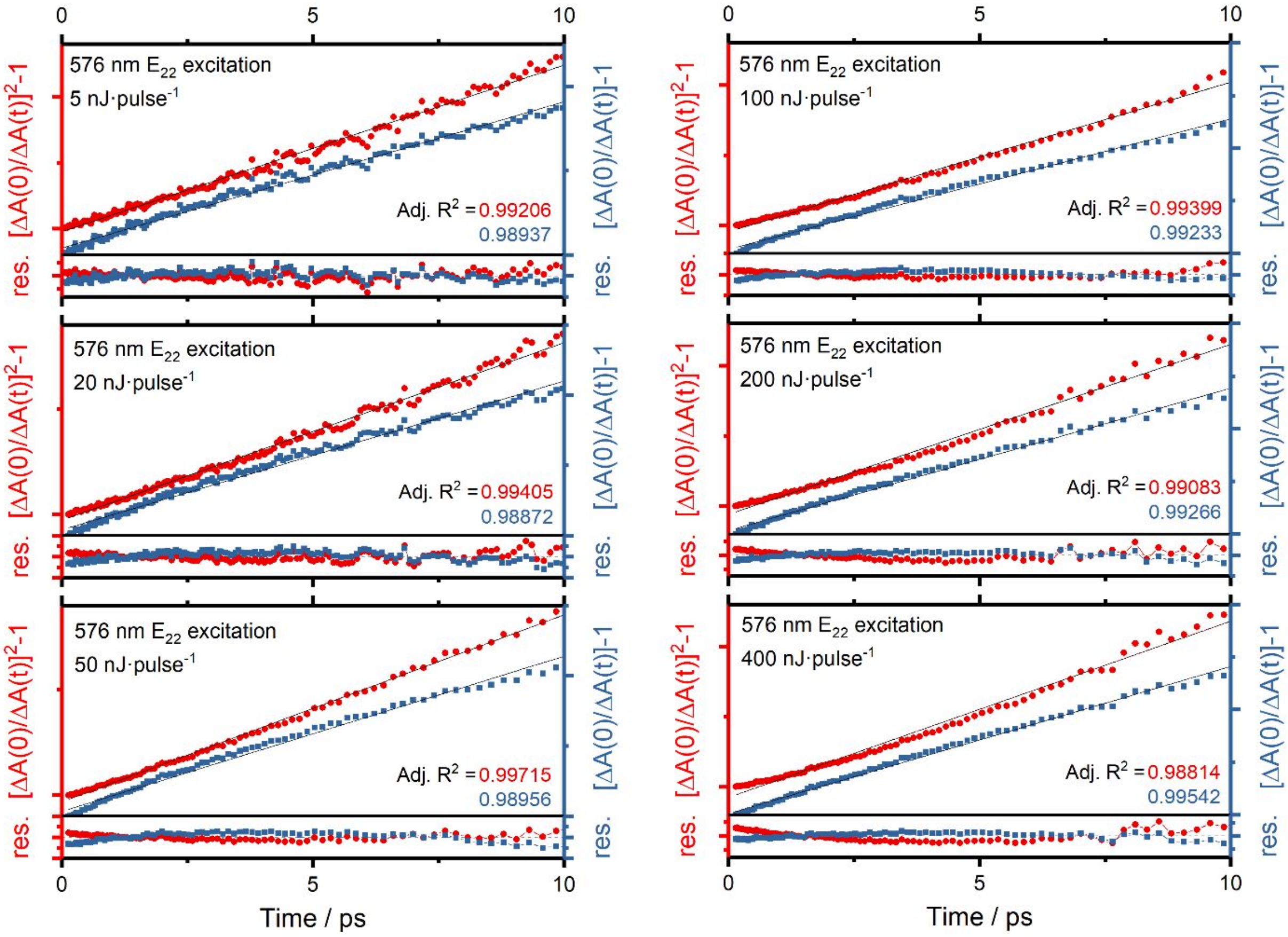

**Figure S17.** Kinetics of the integral $E_{00} \rightarrow E_{11}$ bleaching in the TA spectra of **Hybrid** in THF upon the $E_{22}$ excitations, plotted as $\{[\Delta A(0)/\Delta A(t)]^2-1\}$ (red dots, left axis) and $\{[\Delta A(0)/\Delta A(t)]-1\}$ (blue squares, right axis). Traces are shifted by different offset on the vertical axis for a better comparison. Solid black lines represent the results of the linear fitting. Adjusted R-squared and fitting residuals are shown with corresponding colors. Pump energy: 5, 20, 50, 100, 200, and 400 nJ·pulse$^{-1}$, as indicated in the legends.

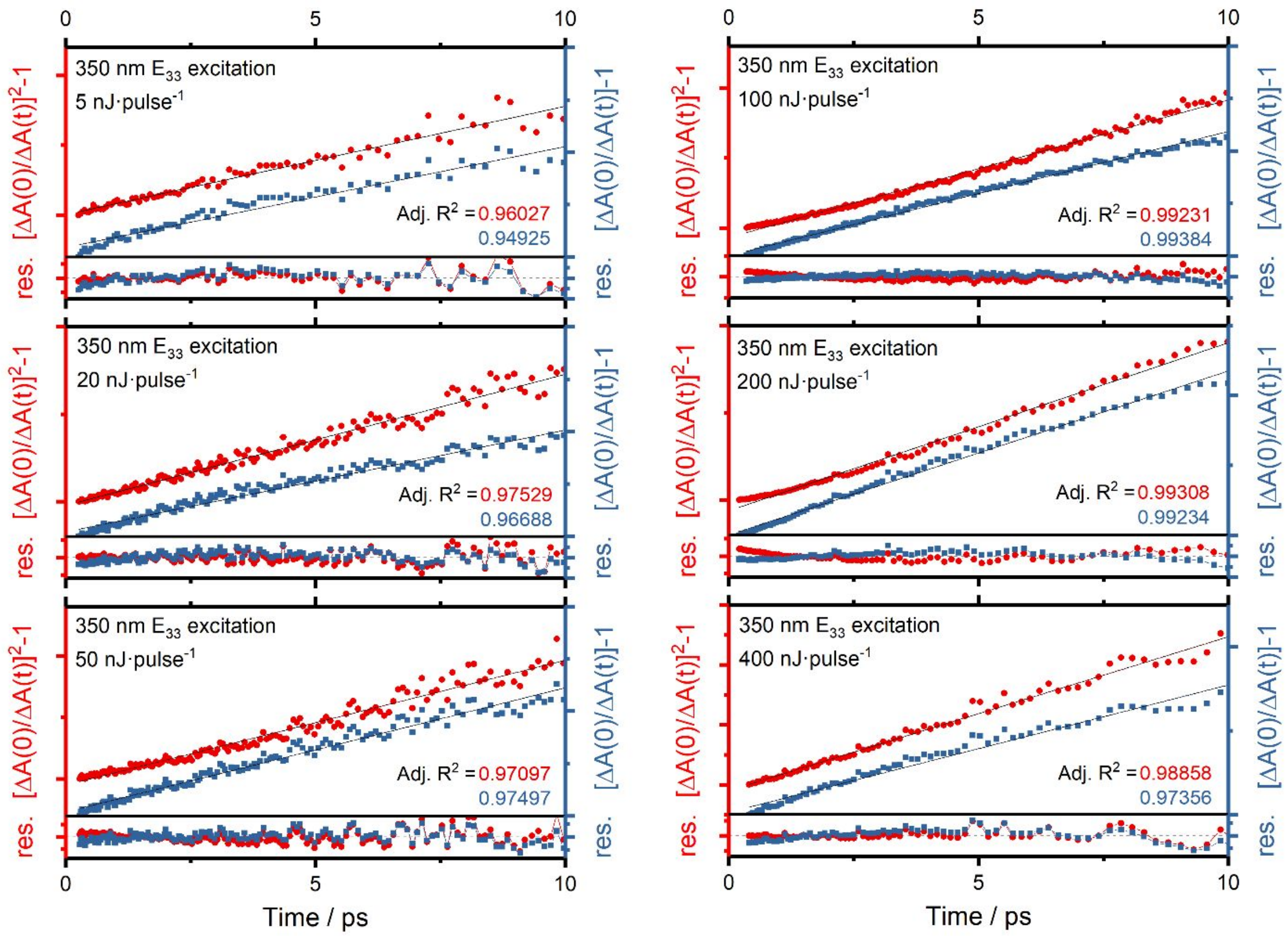

**Figure S18.** Kinetics of the integral $E_{00} \rightarrow E_{11}$ bleaching in the TA spectra of **Hybrid** in THF upon the $E_{33}$ excitations, plotted as $\{[\Delta A(0)/\Delta A(t)]^2-1\}$ (red dots, left axis) and $\{[\Delta A(0)/\Delta A(t)]-1\}$ (blue squares, right axis). Traces are shifted by different offset on the vertical axis for a better comparison. Solid black lines represent the results of the linear fitting. Adjusted R-squared and fitting residuals are shown with corresponding colors. Pump energy: 5, 20, 50, 100, 200, and 400 nJ·pulse$^{-1}$, as indicated in the legends.

**Table S3.** Summary of the Adjusted R-Squared Obtained in Time Dependence Analysis of the $E_{00} \rightarrow E_{11}$ bleaching kinetics of **Hybrid** in THF.*

| excitation wavelength / nm | model | pump energy / nJ·pulse$^{-1}$ | | | | | |
|---|---|---|---|---|---|---|---|
| | | 5 | 20 | 50 | 100 | 200 | 400 |
| 1000 | exciton model | 0.97373 | 0.98727 | 0.99275 | 0.98561 | 0.98311 | -- |
| | carrier model | 0.97093 | 0.98399 | 0.98927 | 0.98016 | 0.98404 | -- |
| 576 | exciton model | 0.98937 | 0.98872 | 0.98956 | 0.99233 | 0.99266 | 0.99542 |
| | carrier model | 0.99206 | 0.99405 | 0.99715 | 0.99399 | 0.99083 | 0.98814 |
| 350 | exciton model | 0.94925 | 0.96688 | 0.97497 | 0.99384 | 0.99234 | 0.97356 |
| | carrier model | 0.96027 | 0.97529 | 0.97097 | 0.99231 | 0.99308 | 0.98858 |

*Combining with the fitting results and residuals shown in Figure S16, S17, and S18, the adjusted R-squared values obtained from the linear fitting are summarized in the table. Exciton model corresponds to the plot of $\{[\Delta A(0)/\Delta A(t)]-1\}$ verse $t$, and the carrier model corresponds to the plot of $\{[\Delta A(0)/\Delta A(t)]^2-1\}$ verse $t$. A larger adjusted R-squared represents a better linear dependence. The larger values of adjusted R-squared in comparison of exciton and carrier models are marked in red.

## H. Analysis of Charge Transfer Dynamics in SWCNT/PFO-BPy Hybrid

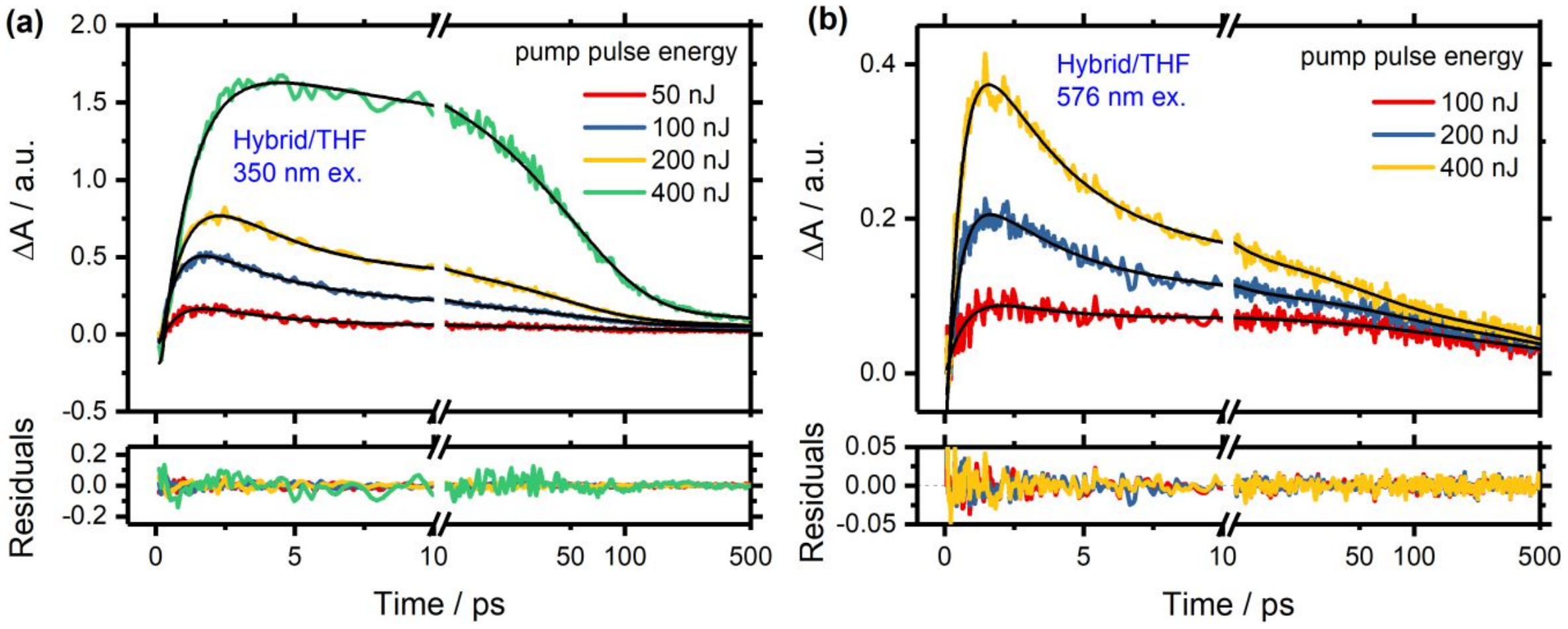

**Figure S19.** TA kinetics trace and fitting results for **Hybrid** in THF at the wavelength of 1050 nm upon (a) 350 nm and (b) 576 nm excitation. Data fits are displayed as black solid lines. Residuals are displayed in the lower panel with corresponding colors. The $E_{11} \rightarrow E_{11,BX}$ absorption is subtracted from these kinetic traces. See Table S1 for the details of fitted parameters.

As discussed in the main text, the TA kinetic traces around 1050 nm of **Hybrid** in THF feature the absorptive signature of SWCNT polaron overlapping with the weak absorption edge of $E_{11} \rightarrow E_{11,BX}$. These kinetic traces are normalized (Figure 7a,b,c in the main text) and subtracted the kinetic traces of 5 nJ, which are regarded as the kinetics of $E_{11} \rightarrow E_{11,BX}$. Thus the obtained kinetic traces represent the pure dynamics of SWCNT polaron (Figure S19). Above traces are adequately fitted with four exponential components, of which one component represents for the rise and three components represent for decays. The fitted parameters are shown in Table S4.

**Table S4.** Transient absorption traces fitting results of **Hybrid** in THF.

| excitation wavelength / nm | pulse energy / nJ | $\tau_1$ /ps ($a_1$)* | $\tau_2$ /ps ($a_2$)* | $\tau_3$ /ps ($a_3$)* | $\tau_4$ /ps ($a_4$)* |
|---|---|---|---|---|---|
| 350 | 50 | 0.69 ± 0.06 (-0.57) | 2.58 ± 0.30 (0.35) | 52.7 ± 25.8 (0.04) | 1377 ± 1289 (0.04) |
| 350 | 100 | 0.62 ± 0.01 (-0.56) | 3.35 ± 0.10 (0.32) | 52.3 ± 3.6 (0.09) | 2143 ± 1117 (0.03) |
| 350 | 200 | 1.04 ± 0.04 (-0.53) | 2.45 ± 0.14 (0.35) | 43.0 ± 1.6 (0.10) | 1019 ± 243 (0.02) |
| 350 | 400 | 1.21 ± 0.15 (-0.55) | 2.42 ± 1.69 (0.09) | 51.0 ± 1.4 (0.33) | 1200 ± 602 (0.03) |
| 576 | 100 | 0.78 ± 0.23 (-0.50) | 1.80 ± 0.89 (0.25) | 102 ± 58 (0.09) | 1178 ± 921 (0.16) |
| 576 | 200 | 0.55 ± 0.02 (-0.55) | 3.43 ± 0.24 (0.28) | 67.8 ± 18.1 (0.07) | 918 ± 286 (0.10) |
| 576 | 400 | 0.59 ± 0.02 (-0.55) | 3.15 ± 0.13 (0.32) | 47.7 ± 6.4 (0.06) | 785 ± 133 (0.06) |

*$\tau_1$, $\tau_2$, $\tau_3$ and $\tau_4$ denote fitted time constants; $a_1$, $a_2$, $a_3$ and $a_4$ denote normalized pre-exponential factors, of which a negative value represents for rise and a positive value represents for decay.

## I. Average Exciton Densities in Transient Absorption Measurements

In transient absorption measurements, we adjusted the concentration of the SWCNTs dispersions to approximated OD = 0.2 at the $E_{11}$ transition in the spectral cuvette with optical length of 2 mm. The number of carbon per unit (atoms/nm) is calculated by $N = \frac{4\Lambda}{3b}$, where $\Lambda = \sqrt[2]{(n^2 + m^2 + mn)}$ with *m* and *n* being the chiral indices of SWCNTs, and *b* being the C-C bond length (nm),[4] thus there are ca. 83.04 C-atoms per nm on the (6,5) SWCNTs. Based on the assumption of linear absorption response, the exciton densities upon certain excitation condition can be estimated. The molar extinction coefficient for the $E_{11}$ absorption of (6,5) per C-atom is 4400 $M^{-1} \cdot L^{-1}$.[5] The mass concentration of the (6,5) SWCNTs is estimated to be $2.73 \times 10^{-6}$ mg/mL. According to the excitation wavelength, spot size, and the extinction coefficients at corresponding wavelengths, the average exciton densities per 100 nm were calculated as shown in Table S5. The average exciton distance on the SWCNTs can be defined as the reciprocal of the exciton linear density. It should be noted that, in reality, EEA prevents such high exciton densities from being reached in SWCNTs and exhibit saturation-like behavior.[6]

**Table S5.** Average exciton densities for the various excitation fluences in transient absorption measurements.

| excitation wavelength / nm | photon energy / eV | pulse energy / nJ | exciton density /100 $nm^{-1}$ | exciton distance / nm |
|---|---|---|---|---|
| 1000 | 1.24 | 5 | 5.74 | 17.4 |
| | | 20 | 23.0 | 4.35 |
| | | 50 | 57.4 | 1.74 |
| | | 100 | 115 | 0.87 |
| | | 200 | 230 | 0.43 |
| | | 400 | 459 | 0.22 |
| 576 | 2.15 | 5 | 1.42 | 70.3 |
| | | 20 | 5.69 | 1.76 |
| | | 50 | 14.2 | 7.03 |
| | | 100 | 28.4 | 3.52 |
| | | 200 | 56.9 | 1.76 |
| | | 400 | 114 | 0.88 |
| 350 | 3.54 | 5 | 8.57 | 117 |
| | | 20 | 3.43 | 29.2 |
| | | 50 | 8.57 | 11.7 |
| | | 100 | 17.1 | 5.84 |
| | | 200 | 34.3 | 2.92 |
| | | 400 | 68.5 | 1.46 |

## J. Comparison between All-Optical Doping and Chemical Doping

To compare the doping level of the SWCNTs between the all-optical excitation and chemical oxidation, we extract the $E_{11}$ peak position as a function of oxidant concentration in the reduction titration experiment (see main text, Figure 6a). As shown in Figure S20, the $E_{11}$ peak progressively blue-shifts with increasing oxidant concentration and reaches a plateau at an oxidant concentration of around 50 μM. As discussed in the main text, Section *Transient Absorption Spectra of SWCNT/PFO-BPy Hybrid*, the dynamic blue-shift of the $E_{11}$ band during the first 50 ps indicates that the share of the SWCNT polaron increases among all quasiparticles. As shown in Figure 5, the $E_{11}$ band could be blue-shifted up to around 989 nm. Thus the all-optical doping of the SWCNT in the Hybrid system could reach a doping level equivalent to chemical oxidation with ~ 8 μM $NOBF_4$. However, because of the spectral overlap in the TA spectra, we could not extract the pure $E_{00} \rightarrow E_{11}$ bleach of pure SWCNT polaron at the delay time that the population of SWCNT polaron reaches its maximum at around 1-3 ps. Therefore, we estimate that the all-optical doping by PCT could reach a doping level higher than the chemical oxidation with ~ 8 μM $NOBF_4$.

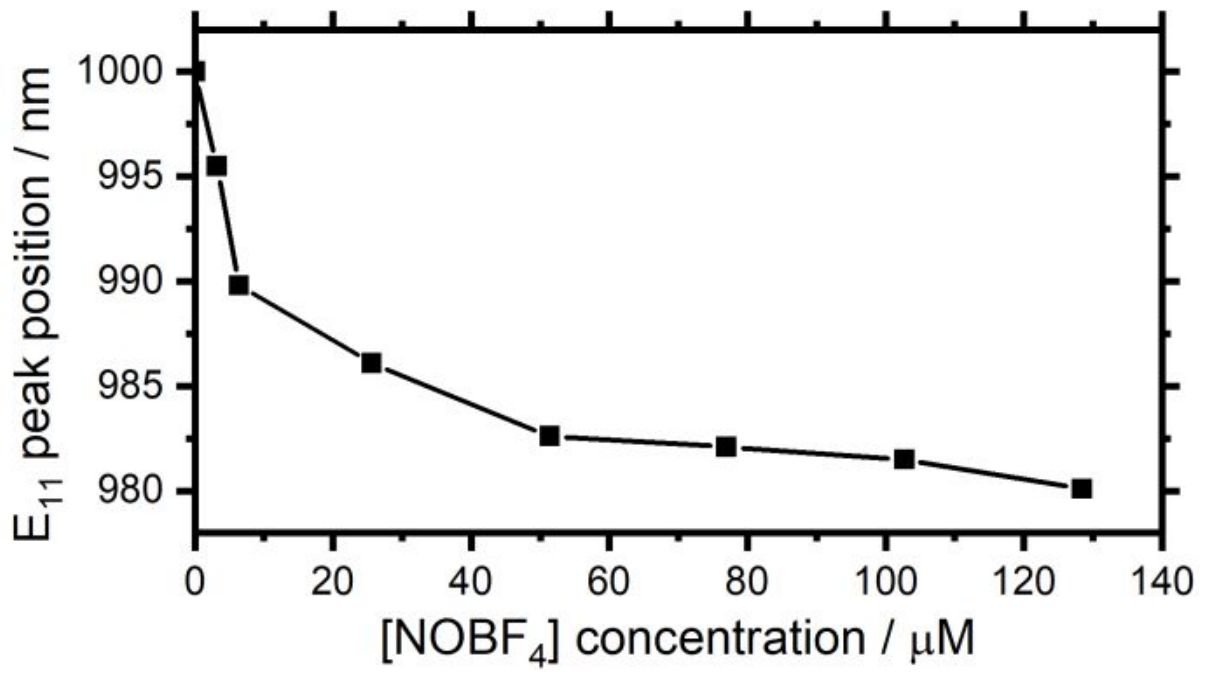

**Figure S20**. $E_{11}$ peak spectral blue-shift observed with increasing oxidant concentration (see main text, Figure 6a) for the **Hybrid** with $NOBF_4$ in toluene:$CH_2Cl_2$ (ratio 1:1) mixed solution.

# References

1. Pfohl, M.; Tune, D. D.; Graf, A.; Zaumseil, J.; Krupke, R.; Flavel, B. S., Fitting Single-Walled Carbon Nanotube Optical Spectra. *ACS Omega* **2017,** *2* (3), 1163-1171.
2. Wei, L.; Flavel, B. S.; Li, W.; Krupke, R.; Chen, Y., Exploring the upper limit of single-walled carbon nanotube purity by multiple-cycle aqueous two-phase separation. *Nanoscale* **2017,** *9* (32), 11640-11646.
3. Eckstein, A.; Karpicz, R.; Augulis, R.; Redeckas, K.; Vengris, M.; Namal, I.; Hertel, T.; Gulbinas, V., Excitation quenching in polyfluorene polymers bound to (6,5) single-wall carbon nanotubes. *Chem. Phys.* **2016,** *467*, 1-5.
4. Byron Pipes, R.; Frankland, S. J. V.; Hubert, P.; Saether, E., Self-consistent properties of carbon nanotubes and hexagonal arrays as composite reinforcements. *Compos. Sci. Technol.* **2003,** *63* (10), 1349-1358.
5. Schöppler, F.; Mann, C.; Hain, T. C.; Neubauer, F. M.; Privitera, G.; Bonaccorso, F.; Chu, D.; Ferrari, A. C.; Hertel, T., Molar Extinction Coefficient of Single-Wall Carbon Nanotubes. *J. Phys. Chem. C* **2011,** *115* (30), 14682-14686.
6. Schneck, J. R.; Walsh, A. G.; Green, A. A.; Hersam, M. C.; Ziegler, L. D.; Swan, A. K., Electron Correlation Effects on the Femtosecond Dephasing Dynamics of E22 Excitons in (6,5) Carbon Nanotubes. *J. Phys. Chem. A* **2011,** *115* (16), 3917-3923.